\documentclass[prd,preprint,singlecolumn,superscriptaddress,preprintnumbers,amsmath,amssymb,tightenlines,longbibliography,nofootinbib]{revtex4-2}

\usepackage{epsfig}  
\usepackage{graphicx}
\usepackage{color}
\usepackage[dvipsnames]{xcolor}
\usepackage{float}
\usepackage{amsfonts}
\usepackage{amsmath}
\DeclareMathOperator{\sign}{sgn}
\usepackage[compat=1.0.0]{tikz-feynman}
\usepackage{slashed}
\usepackage{mathrsfs}
\usepackage{slashed}
\usepackage{soul}
\usepackage{braket}
\usepackage{dsfont}
\usepackage[colorlinks,citecolor=blue,urlcolor=blue,bookmarks=false,hypertexnames=true]{hyperref}
\usepackage{orcidlink}

\large

\begin{document} 

\title{Probing GPDs in exclusive electroproduction of dijets}

\author{Trambak Jyoti Chall\orcidlink{000000025932278X}}
\email{trambakjyotic@dokt.ur.edu.pl}
\affiliation{Doctoral School, Institute of Physical Sciences, University of Rzeszów, Al. Tadeusza Rejtana 16C, PL-35-959 Rzeszów, Poland}

\author{Marta Łuszczak\orcidlink{0000000317768019}}
\email{mluszczak@ur.edu.pl}
\affiliation{Institute of Physical Sciences, University of Rzeszów, Al. Tadeusza Rejtana 16C, PL-35-959 Rzeszów, Poland}

\author{Wolfgang Schäfer\orcidlink{0000000207640372}}
\email{wolfgang.schafer@ifj.edu.pl}
\affiliation{Institute of Nuclear Physics, Polish Academy of Sciences, ul. Radzikowskiego 152, PL-31-342 Kraków, Poland}

\author{Antoni Szczurek\orcidlink{0000000152478442}}
\email{antoni.szczurek@ifj.edu.pl}
\affiliation{Institute of Nuclear Physics, Polish Academy of Sciences, ul. Radzikowskiego 152, PL-31-342 Kraków, Poland}

\begin{abstract}
We summarize the formalism for calculating the exclusive dijet production
in $e p \to e^{\prime} jj p$ in collinear QCD factorization, using generalized parton distributions as the soft hadronic input modeled in the double distribution approach. 
We include all leading-order contributions coming from light sea and valence quark exchanges, and gluon exchanges for both light quark-antiquark ($q\bar{q}$) production and also the heavy $c\bar{c}$ final state.
We present results for several differential distributions for the cross section evaluated over a broad region of phase space, covering a wide range of inelasticity and photon virtuality.
The gluon and sea contributions exhibit similar shapes, whereas the valence contribution, though relatively small, shows a markedly different behavior. The latter becomes particularly noticeable at large $x_{\mathbb{P}}$, a kinematic region not explored at HERA, but potentially accessible in future measurements at the Electron Ion Collider.
This requires further feasibility studies.
We also present the azimuthal angle modulation between the leptonic and the outgoing dijet planes for the general case, as well as for the ZEUS kinematic region where we see reasonable agreement with the data for diffractive deep inelastic scattering parameter $\beta \gtrsim 0.4$.
\end{abstract}

\maketitle
\newpage

\section{Introduction}

Diffractive deep inelastic scattering (DIS) has traditionally been a testbed for ideas on the QCD Pomeron \cite{Nikolaev:1991et,Barone:2002cv}.
Diffractive events are generally defined by large rapidity gaps in the hadronic final state, often with the proton left intact.
The diffractive production of exclusive and inclusive dijets was studied using different formalisms \cite{Nikolaev:1994cd,Bartels:1996tc,Golec-Biernat:1998exl,Bartels:1999tn,Braun:2005rg, Altinoluk:2015dpi,Britzger:2018zvv,Salazar:2019ncp,Mantysaari:2019csc}.
Recently, in \cite{Linek:2024dzs}, a Generalized Transverse Momentum Distribution (GTMD) formalism, which is closely related to the color dipole approach \cite{Nikolaev:1990ja,Nikolaev:1991et}, was used to calculate numerous differential distributions in HERA kinematics.
There, it was shown that it is not possible to describe the diffractive dijet production process of both H1 \cite{H1:2011kou} and ZEUS \cite{ZEUS:2015sns} data simultaneously by the production of exclusive $q \bar q$ dijet systems.
In most of the approaches available in the literature, only gluon degrees of freedom were considered for the color-singlet $t$--channel exchange.

A notable exception in this respect is the collinear factorization approach of \cite{Braun:2005rg} based on 
generalized parton distributions (GPDs). GPDs can be probed in exclusive scattering processes where they encode the internal three-dimensional structure of hadrons in terms of both quark and gluon degrees of freedom, see for example the reviews \cite{Ji:1998pc,Diehl:2003ny,Belitsky:2005qn} . 
In the GPD approach also  contributions to the exclusive dijet production corresponding to sea and valence quark exchanges have been considered.
One of the primary aims of this work is to segregate the contributions to the differential cross section across sea and valence quark exchanges and also gluon exchanges for both light $q\bar{q}$ and heavy $c\bar{c}$ final state. We show these components in an extensive phase space across a variety of kinematical variables so that we can understand and analyze the behavior of each of the contributions coming from the various partonic exchanges in a broad kinematic range, not just limited to the purely diffractive, large rapidity gap, region. 

To our knowledge, the collinear factorization GPD approach was never discussed in the context of the HERA data \cite{H1:2011kou,ZEUS:2015sns}. Therefore, we intend to do it here for the first time. We take the kinematical region corresponding to HERA by imposing the relevant cuts and analyze the region of agreement with the available data. 
Since there are no commonly available implementations of the formalism, we had to reproduce the calculations of Braun and Ivanov \cite{Braun:2005rg} in order to develop our own framework for the analysis.
In this work, we discuss the most generic situation for exclusive dijet production, first without any experimental constraint, and then we also show the relevant results corresponding to the kinematics at the ZEUS \cite{ZEUS:2015sns} experiment at HERA. 

The work is organized as follows: 
the next section, Sec.~\ref{sec:2} on theoretical formalism is divided into four parts, in the first part Sec.~\ref{sec:2a}, we set up the kinematics for the dijet production process, in the second part Sec.~\ref{sec:2b}, we introduce the GPDs and then explicitly show the calculation of the amplitude of the $\gamma^{\ast}p$ scattering for both longitudinal and transverse photon polarizations, in the third part Sec.~\ref{sec:2c}, we show the cross section of interest and in the fourth part Sec.~\ref{sec:2d}, we show the modeling of the GPDs that we used in the double distribution (DD) approach.
The third section Sec.~\ref{sec:3} on results and discussions is divided into two parts, in the first part, Sec.~\ref{sec:3a}, we show our results for an extensive phase space, and in the next part, Sec.~\ref{sec:3b}, we show our results corresponding to ZEUS kinematics. We provide our concluding remarks in the final section, Sec.~\ref{sec:4}.

\section{Theoretical Formalism}
\label{sec:2}
\subsection{Kinematics}
\label{sec:2a}
\begin{figure*}[htb!]
    \centering   
    \includegraphics[width=0.65\linewidth]{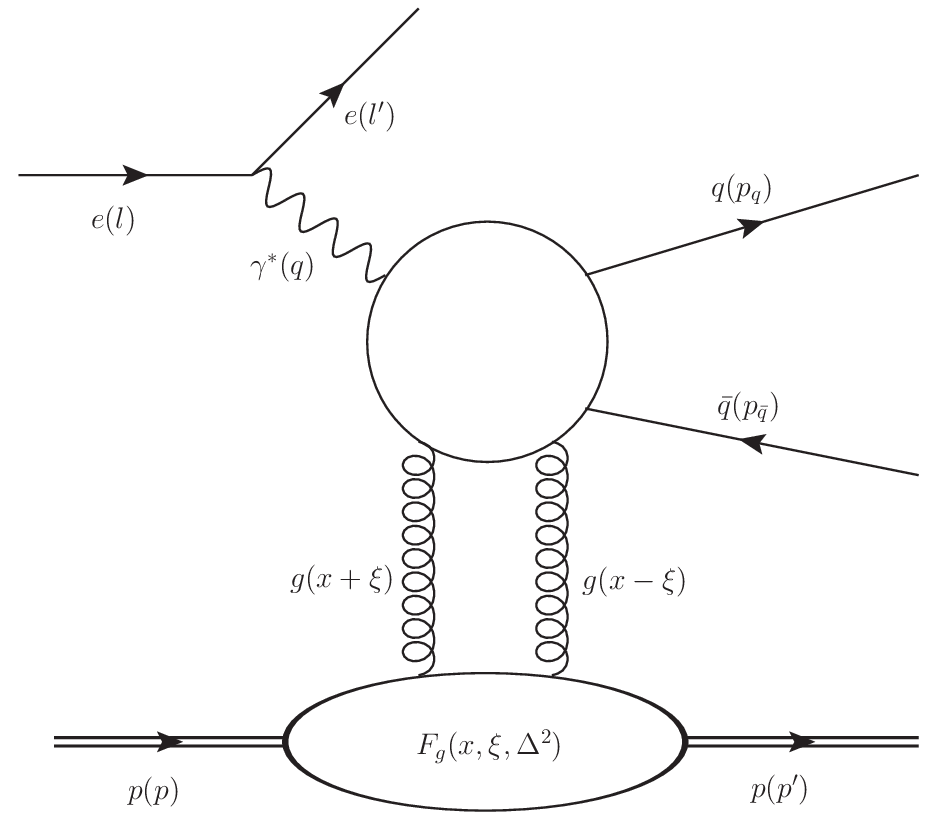}
    \caption{The hard and exclusive diffractive dijet (jj) production in deep inelastic electron-proton scattering in the two-gluon exchange model.}
    \label{fig:plot1}
\end{figure*}

We start by introducing the standard kinematical variables of diffractive DIS. We denote the lepton and four-momenta by $l, l'$, and the four momenta of incoming and outgoing protons by $p, p'$ respectively, see Fig.~\ref{fig:plot1}. 
The four-momentum carried by the virtual photon is $q = l-l'$, and its virtuality is $Q^2 = - q^2$.

The commonly used variables are the $ep$ center-of-mass energy squared $s = (l+p)^2 \approx  2 \,  l\cdot p$, as well as the  $\gamma^* p$ center-of-mass energy squared $W^2 = (q+p)^2 \approx 2 q \cdot p - Q^2$.
As we are interested in the deep--inelastic regime $W^2, Q^2 \gg m_p^2$, as indicated, we will neglect the proton mass wherever possible, as we do for the leptons.

Furthermore, we make use of the Bjorken variable 
\begin{eqnarray}
    x_{\rm Bj} = \frac{Q^2}{2 p \cdot q} = \frac{Q^2}{Q^2+W^2} \, , 
\end{eqnarray}
and the inelasticity
\begin{eqnarray}
y = \frac{ q \cdot P }{ l \cdot P} = \frac{W^2 + Q^2}{s} = \frac{Q^2}{x_{\rm Bj}s} \, . 
\end{eqnarray}
We are interested in diffractive processes, in which the proton stays intact after receiving a four momentum transfer $\Delta = p'-p$ which defines the Mandelstam variable $t = \Delta^2$ and a diffractive system of invariant mass $M^2 = (q + p - p')^2 = (q - \Delta)^2$ is produced.

The counterpart of $y$ for the $p \to p'$ vertex is
\begin{eqnarray}
x_{\mathbb{P}} = \frac{\Delta \cdot q}{p \cdot q} = \frac{ M^2 + Q^2 - t}{W^2 + Q^2} \approx \frac{ M^2 + Q^2}{W^2 + Q^2},
\end{eqnarray}
and we also use\footnote{Here the superscript ``DDIS'' on the $\beta$ parameter stands for diffractive DIS.} 
\begin{eqnarray}
\beta^{\rm DDIS} = \frac{Q^2}{2 \Delta \cdot p} = \frac{Q^2}{M^2 + Q^2 - t} = \frac{x_{\rm Bj}}{x_{\mathbb P}} \approx \frac{Q^2}{M^2 + Q^2}\,.
\end{eqnarray}

The limit of a nearly lightlike incident momenta of the colliding proton, i.e. $p^2\rightarrow{0}$, motivates the introduction of two lightlike directions
\begin{equation}
p_\mu , \quad q_{\mu}^{\prime}=q_{\mu}+x_{\rm Bj}p_{\mu}\,, \quad p^2 = q'^2 = 0. 
\end{equation}
We adopt a coordinate system in which the $\gamma^* p$ collision axis is along the $z$ direction.
Lepton four-momenta then have the form
\begin{eqnarray}
 l_\mu = \frac{1-y}{y} x_{\rm Bj} p_\mu  + \frac{1}{y} q'_\mu + l_{\perp \mu}, \nonumber \\
 l'_\mu = \frac{1}{y} x_{\rm Bj} p_\mu  + \frac{1-y}{y} q'_\mu + l_{\perp \mu},  
\end{eqnarray}
with a lepton transverse momentum 
\begin{eqnarray}
    \vec l_\perp^{~2} = - l_\perp^2 = \frac{1-y}{y^2} \, Q^2\,.
\end{eqnarray}
In the following we follow closely the notation and conventions of \citet{Braun:2005rg}.
Introducing the lightlike vectors
\begin{eqnarray}
    n_\pm = \frac{1}{\sqrt{2}} (1, 0, 0, \pm 1), \quad n_+ \cdot n_- = 1, 
\end{eqnarray}
we write
\begin{eqnarray}
    p = W(1+ \xi) n_+ , \quad q' = \frac{W^2 + Q^2}{2W (1+ \xi)} n_-,
\end{eqnarray}
so that the photon carries dominant light cone minus momentum
\begin{eqnarray}
 q = -W (1+ \xi) x_{\rm Bj} n_+ + \frac{Q^2 + W^2}{2W (1+ \xi)} n_-,
\end{eqnarray}
and the outgoing proton momentum is
\begin{eqnarray}
p' = W(1-\xi) n_+ + \frac{\Delta_\perp^2}{2 W(1-\xi)} n_- + \Delta_\perp .
\end{eqnarray}
Here we have introduced the skewness or asymmetry parameter $\xi$ given by
\begin{eqnarray}
    \xi = \frac{ (p - p') \cdot n_-}{(p+p')\cdot n_-}, \quad \frac{2 \xi}{1+\xi} = \frac{M^2 + Q^2}{W^2+Q^2} = x_{\mathbb P}.
\end{eqnarray}
Below, we also use the average of incoming and outgoing proton momenta
\begin{eqnarray}
P = \frac{1}{2} ( p + p') .
\end{eqnarray}

The four-momenta of quark and antiquark jet of flavor $f$ are, respectively
\begin{eqnarray}
p_q &=& z q' + \frac{
\vec p_\perp^{~2} + m_f^2}{2 z \, q'\cdot n_+} n_+  + p_\perp, \nonumber \\
p_{\bar q} &=& (1-z) q' + \frac{
\vec p_\perp^{~2} + m_f^2}{2 (1-z) \, q'\cdot n_+} n_+  -  p_\perp\,,
\end{eqnarray}
where $p_{\perp}$ is the jet transverse momentum.
Notice that we treat the dijets in back-to-back kinematics, neglecting the momentum transfer $\Delta_\perp$ in the hard matrix element.
The invariant mass of the dijet system then becomes
\begin{eqnarray}
   M^2 = (p_q + p_{\bar q})^2 = \frac{\vec p_\perp^{~2} + m_f^2}{z(1-z)}.
\end{eqnarray}
The asymmetry parameter can be expressed in terms of the aforementioned quantities as
\begin{equation}
    \xi=\frac{x_{\mathbb{P}}}{2-x_{\mathbb{P}}}\,,
\end{equation}
where $x_{\mathbb{P}}=x_{\rm Bj}/\beta$ is the momentum fraction of the proton carried by the so-called ``pomeron'' corresponding to the gluon ladder shown in Fig.\ref{fig:plot1}. 
Following \citet{Braun:2005rg}, we define a $\beta$ parameter as\footnote{This is our definition of $\beta$ throughout the work except in Sec.~\ref{sec:2d}, where $\beta$ is used to represent the fraction of momenta carried away by the outgoing parton from its parent hadron.},
\begin{equation}
    \beta= \frac{\varepsilon^2}{\vec{p}_\perp^{~2} + \varepsilon^2} = \beta^{\rm DDIS}\left(1+\frac{m_f^2}{Q^2 z (1-z)}\right)\,, \quad \varepsilon^2 \equiv m_f^2 + z(1-z)Q^2,
    \label{beta}
\end{equation}
which for $m_f \to 0$  coincides with the standard  diffractive DIS variable $\beta^{\rm DDIS}$.
In this work, we have always considered $\beta=\beta^{\rm DDIS}$ except for the case of the heavy $c\bar{c}$ production through gluon exchanges, where the charm mass has to be weighed into the definition of the $\beta$ parameter as per the above relation.
The rapidities of both the outgoing jets in the HERA frame are expressed as
\begin{equation}
    \eta_1=-\ln\left(\frac{zW^2}{2p_{\perp} E_p}\right)\,,
    \quad \eta_2=-\ln\left(\frac{(1-z)W^2}{2p_{\perp} E_p}\right)\,,
    \label{eta1and2}
\end{equation}
where $E_p=920\,\rm GeV$ is the energy of the proton beam.
From the above Eq.~\eqref{eta1and2}, we can clearly see that $\Delta\eta=|\eta_1 - \eta_2|=|\ln
( (1-z)/z )|$.
Finally, the Mandelstam-$t$ variable can be expressed as
\begin{equation}
    t=- \frac{\vec \Delta_\perp^2+x_{\mathbb{P}}^2m_N^2}{1-x_{\mathbb{P}}} \approx - \frac{\vec \Delta_\perp^2+ 4 \xi^2m_N^2}{1-\xi^2}\,,
    \label{tmand}
\end{equation}
where we restored the finite proton mass $m_N$ in order to put in evidence the minimal value of $|t|$ at $\vec \Delta_\perp = 0$.
\subsection{Calculation of amplitudes}
\label{sec:2b}
\begin{figure*}
    \centering   
    \includegraphics[width=0.48\linewidth]{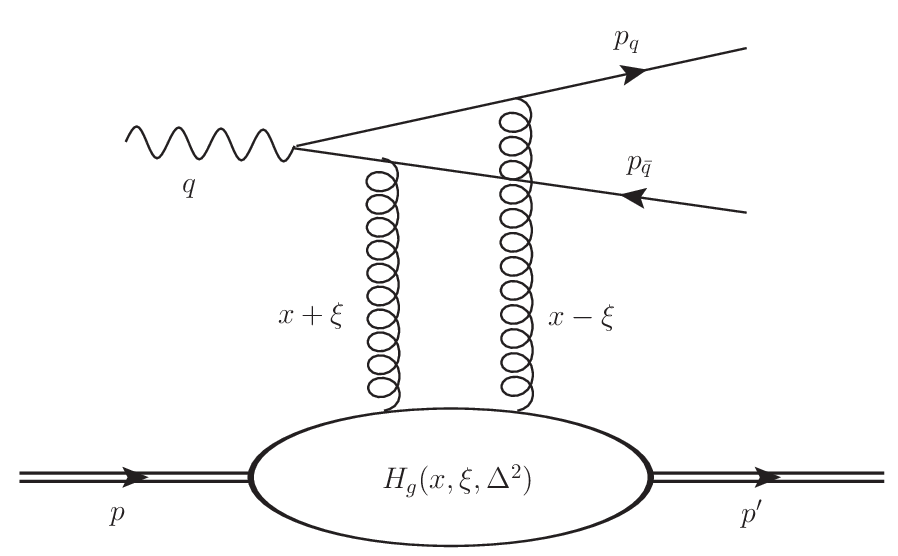} \hspace{1em}
    \includegraphics[width=0.48\linewidth]{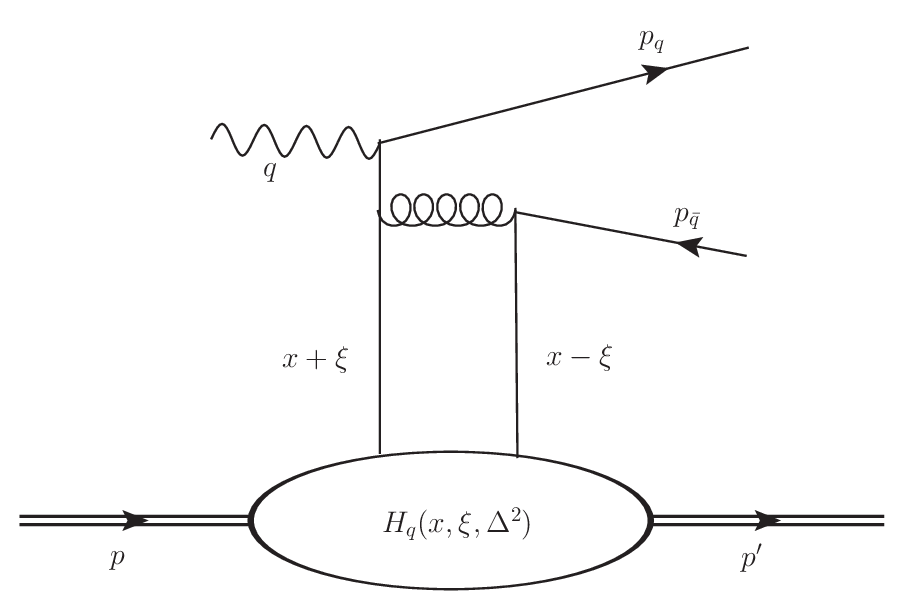}
    \caption{Feynman diagrams contributing to the exclusive dijet production in the $\gamma^{\ast}p\xrightarrow{}q\bar{q}p$ subprocess. The left panel presents the gluon exchange contribution whereas the right panel shows the quark exchange contribution.}
    \label{fig:plot2}
\end{figure*}

In the theoretical framework of collinear QCD factorization, the interaction amplitudes can be expressed as a convolution of GPDs with functions that are perturbatively calculable. 
The total contribution can be divided into ones coming from gluon exchanges shown in the left panel of Fig.~\ref{fig:plot2} and in the right panel the contribution coming from quark exchanges is shown\footnote{Only the first diagram with gluon exchanges contribute to the the heavy $c\bar{c}$ final state.}. 
Here, the GPDs parametrize the hadronic input and can be defined analogously to the usual parton densities as the matrix elements of the light cone bilocal gluon and quark operators sandwiched between the outgoing and incoming hadronic momenta states as \cite{Ji:1998pc,Diehl:2003ny,Belitsky:2005qn},
\begin{eqnarray}
F^g\left(x,\xi,\Delta^2\right)&=&\frac{1}{P \cdot n_{-}} \int \frac{d\lambda}{2\pi} e^{ix(P\cdot u)}n_{-\alpha}n_{-\beta}\bra{p'}G^{\alpha\mu}
    \left(-\frac{u}{2}\right)G^{\beta}_{\mu}
    \left(\frac{u}{2}\right)\ket{p}\Bigg|_{u=\lambda n_{-}}\,,\nonumber\\F^q\left(x,\xi,\Delta^2\right)&=&\frac{1}{2} \int \frac{d\lambda}{2\pi} e^{ix(P \cdot u)}\bra{p'}\bar{q}\left(-\frac{u}{2}\right)\slashed{n}_{-} q\left(\frac{u}{2}\right)\ket{p}\Bigg|_{u=\lambda n_{-}}\,,
\end{eqnarray}
wherein we are integrating with a Fourier phase $e^{ix(P\cdot u)}$, the bilocal matrix elements of gluon fields (represented by the vectors $G^{\alpha}$ and $G^{\beta}$) and quark fields (represented by $q$ and $\bar{q}$) separated along the lightlike direction of $n_{-}$. 

In these Fourier integrals, the Lorentz decomposition of the matrix elements corresponding to the two possible Dirac structures (allowed by Lorentz symmetry and parity)--- $\bar{u}(p')\slashed{n}_{-}u(p)$ and $\bar{u}(p')i\sigma^{\alpha\beta}n_{-\alpha}\Delta_{\beta}u(p)$ allows us to project out the GPDs $H^{g,q}(x,\xi, \Delta^2)$ and $E^{g,q}(x,\xi, \Delta^2)$ respectively as scalar coefficients of this expansion as
\begin{equation}
F^{g,q}\left(x,\xi,\Delta^2\right)=\frac{1}{2(P \cdot n_{-})}\left[H^{g,q}(x,\xi,\Delta^2)\bar{u}(p')\slashed{n}_{-}u(p)+E^{g,q}(x,\xi,\Delta^2)\bar{u}(p')\frac{i\sigma^{\alpha\beta}n_{-\alpha}\Delta_{\beta}}{2m_N}u(p)\right]\,,
    \label{HE}
\end{equation}
where $m_N$ denotes the nucleon mass, and $u(p)$ and $\bar{u}(p')$ represent the nucleon spinors. In this work, we are interested only in the helicity--conserving contribution, as it dominates in the region of small $|\vec \Delta_\perp|$ within the so-called diffraction cone.
The helicity nonflip contribution picks out \cite{Diehl:2003ny}
\begin{eqnarray}
    F^{g,q}(x,\xi,\Delta^2) \propto H^{g,q}(x,\xi,\Delta^2) - \frac{\xi^2}{1-\xi^2}  E^{g,q}(x,\xi,\Delta^2). 
\end{eqnarray}
As we are mainly interested in HERA data taken at small $x_{\mathbb P}$, and hence small $\xi$, we also neglect the contribution of GPDs $E^{g,q}$ in the following. Their inclusion into the presented formalism is however straightforward.

Also assuming equal helicities of the initial and final state hadrons, the skewness or asymmetry parameter $\xi=0$ and the GPDs $H^{g,q}(x,0,0)$ reduce to the ordinary spin-independent nucleon densities as\footnote{Here we are only dealing with unpolarized GPDs $H^{g,q}$ and $E^{g,q}$. One can, in principle, also consider the spin-polarized GPDs $\tilde{H}^{g,q}$ and $\tilde{E}^{g,q}$.},
\begin{eqnarray}
    H^g(x,0,0)&=&x\left[\theta(x)g(x)+\theta(-x){g}(-x)\right]\,,\nonumber\\ H^q(x,0,0)&=&\theta(x)q(x)-\theta(-x)\bar{q}(-x)\,,
\end{eqnarray}
where $g(x)$ and $q(x)$ are the usual gluon and quark parton distribution functions (PDFs). We can notice here that unlike $H^q$, since gluons are their own antiparticles, $H^g$ is an even function in $x$.

The amplitude for the subprocess of virtual photon-proton $\gamma^{\ast}p$ scattering for longitudinal photon polarization was obtained in \cite{Braun:2005rg} as
\begin{equation}
    \mathcal{A}_{\gamma^* L}=-\frac{4\pi }{N_c} \alpha_s \sqrt{4\pi\alpha_{em}}e_q \delta_{ij}\frac{z (1-z) QW}{\left[ \vec p_{\perp}^{~2} + \varepsilon^2 \right]^2}\bar{u}(p_q)\slashed{n}_{+}v(p_{\bar q})\left(\mathcal{I}_L^g + 2 C_F \mathcal{I}_L^q \right)\,,
    \label{longA}
\end{equation}
where $\alpha_{em}$ is the electromagnetic and $\alpha_{s}$ is the strong coupling, $e_q$ is the fractional quark charge, $u(p_q)$ and $v(p_{\bar q})$ denote the quark and antiquark spinors, $\delta_{ij}$ is for the color indices of the outgoing quarks, and we have used the QCD color factor $C_F=(N_c^2-1)/(2N_c)$.
 Here, in Eq.~\eqref{longA} the gluon and quark integrals $\mathcal{I}_L^g$ and $\mathcal{I}_L^q$ can be expressed as integrals over $-1\leq x\leq 1$ in the following way \cite{Braun:2005rg}:
\begin{eqnarray}
    \mathcal{I}_L^g &=& \int_{-1}^{1} dx F^g\left(x,\xi,\Delta^2\right)\left( \frac{2\xi (1-\beta)}{\left(x+\xi-i\epsilon \right)^2}+ \frac{2\xi (1- \beta)}{\left(x-\xi+i\epsilon \right)^2}- \frac{2\xi(1-2\beta)}{\left(x+\xi-i\epsilon \right)\left(x-\xi+i\epsilon \right)}\right)\,,\nonumber\\
    \mathcal{I}_L^q &=&\int_{-1}^{1} dx F^q\left(x,\xi,\Delta^2\right)\left( \frac{2\xi (1-z)}{x+\xi-i\epsilon}+ \frac{2\xi z}{x-\xi+i\epsilon}\right)\,,
    \label{IL}
\end{eqnarray}
where a limit $\epsilon\rightarrow{0^+}$ is understood. 

These expressions allow us to decompose the amplitude into analogs of the ``Compton form factors."
For example, the integral ${\cal I}^q_L$ can be written as
\begin{eqnarray}
   \mathcal{I}_L^q &=&\Big\{ zf_q^{(1)}(\xi,\Delta^2)+(1-z) f_q^{(2)}\left(\xi,\Delta^2\right)\Big\} \frac{\bar u(p') \slashed{n}_{-}u(p)}{2(P \cdot n_{-})}, 
\end{eqnarray}
with\footnote{The expressions for these quark integrals shorthand notation can be found in Eq.~\eqref{a6} of the Appendix \ref{sec:Integrals}.}
\begin{eqnarray}
      f_q^{(1)}(\xi,\Delta^2) &=& 2\xi \int_{-1}^{1} dx \, \frac{H^q(x,\xi,\Delta^2)}{x - \xi + i \epsilon },    \quad
    f_q^{(2)}(\xi,\Delta^2) = 2\xi \int_{-1}^{1} dx \, \frac{H^q(x,\xi,\Delta^2)}{x + \xi - i \epsilon } .
\end{eqnarray}
In practice, it is even more convenient to use 
\begin{eqnarray} 
f_q^{\pm}(\xi,\Delta^2) = \xi \int_{-1}^1 dx H^q(x,\xi,\Delta^2) \Big( \frac{1}{x - \xi + i \epsilon } \mp  \frac{1}{x + \xi - i \epsilon } \Big) .
\end{eqnarray}
As the odd and even in $x$ combinations of quark-GPDs
\begin{eqnarray}
    H^{q+}(x,\xi,\Delta^2) &\equiv&  H^{q}(x,\xi,\Delta^2) - H^{q}(-x,\xi,\Delta^2), \nonumber \\
    H^{q-}(x,\xi,\Delta^2) &\equiv&  H^{q}(x,\xi,\Delta^2) + H^{q}(-x,\xi,\Delta^2)\,,
    \end{eqnarray}
correspond to positive and negative charge-conjugation parity, the quark contribution to the longitudinal amplitude can be conveniently decomposed into its $C$ even and $C$ odd parts:
\begin{eqnarray}
  \mathcal{I}_L^q &=&\Big\{ f_q^{+}(\xi,\Delta^2)+(1-2z) f_q^{-}\left(\xi,\Delta^2\right)\Big\} \frac{\bar u(p') \slashed{n}_{-}u(p)}{2(P \cdot n_{-})}\,. 
\end{eqnarray}
Obviously, the interference between the $C$ even and $C$ odd exchanges gives rise to the contribution to the cross section $\propto (1-2z)$ which vanishes after symmetrically integrating over $z$.
Let us remind the reader that gluon and sea quark exchange contributions are $C$--even, while the valence quark contributions contain both charge conjugation parities \cite{Diehl:2003ny}.

The singularities at $x = \pm \xi$ lead to the integrals/Compton form factors having imaginary parts.
The first-order poles as in the quark Compton form factors are easily dealt with.
We can simply apply the Sokhotski-Plemelj formula for real line integrals for first-order poles, given as
\begin{equation}
\lim_{\epsilon\rightarrow{0^+}}\int_{-1}^{1} dx \frac{f(x)}{x\mp \xi\pm i\epsilon }=PV\int_{-1}^{1}dx \frac{f(x)}{x\mp \xi}\mp i\pi f\left(\pm \xi\right)\, , 
\label{SP1}
\end{equation}
and the Cauchy principle value is evaluated from
\begin{eqnarray}
    PV\int_{-1}^{1}dx \frac{f(x)}{x\mp \xi}&=&\int_{-1}^{1}dx\frac{f(x)-f(\pm\xi)}{x\mp\xi}+f\left(\pm\xi\right)\log\left|\frac{1\mp\xi}{1\pm\xi}\right|\,.
    \label{reg}
\end{eqnarray}
However, for the gluon contribution, one also encounters second-order poles.
We deal with them by using
\begin{equation}
\frac{1}{(x \mp \xi \pm i \epsilon)^2} = - \frac{d}{dx} \Big(\frac{1}{x \mp \xi \pm i\epsilon}\Big)\,,
\end{equation}
so that when the boundary terms drop out on the assumption that the integrand function $f(x)$ vanishes there, we are left with,
\begin{eqnarray}
    \int_{-1}^{1} dx  \frac{f(x)}{\left(x\mp \xi\pm i\epsilon \right)^2} = \int_{-1}^{1} dx \frac{f'(x)}{x\mp \xi\pm i\epsilon }. 
\end{eqnarray}
This integral can then be handled by the standard method as above, provided the derivative of the relevant GPD in $x$ is continuous. This is at least the case for the GPDs that we use in this work (see Sec. \ref{sec:2d} below). Otherwise, the appearance of double poles would signal a breakdown of the adopted collinear QCD factorization formalism.

We now move on to the amplitude for transversely polarized photons, obtained in \cite{Braun:2005rg} as:

\begin{eqnarray}
    \mathcal{A}_{\gamma^* T}&=&-\frac{2 \pi}{N_c}\alpha_s \sqrt{4\pi\alpha_{em}}e_q \delta_{ij}\frac{W}{\left[ \vec p_{\perp}^{~2} + \varepsilon^2 \right]^2}\Big\{-\bar{u}(p_q)\left(m\slashed{e}_{\perp}\right)\slashed{n}_{+}v(p_{\bar q})\mathcal{I}_L^g \nonumber\\&+& \bar{u}(p_q)\slashed{p}_{\perp}\slashed{e}_{\perp}\slashed{n}_{+}v(p_{\bar q}) \left((1-z) \mathcal{I}_T^g + 2C_F \mathcal{I}_T^{q_1} \right) \nonumber \\
    &+& \bar{u}(p_q)\slashed{e}_{\perp}\slashed{p}_{\perp}\slashed{n}_{+}v(p_{\bar q}) \left(-z\mathcal{I}_T^g + 2 C_F \mathcal{I}_T^{q_2} \right)\Big\}\,.
    \label{transA}
\end{eqnarray}
Here, the relevant integrals over GPDs are:
\begin{eqnarray}
    \mathcal{I}_T^g &=& \int_{-1}^{1} dx F^g\left(x,\xi,\Delta^2\right)\left( \frac{\xi\left(1-2\beta\right)}{\left(x+\xi-i\epsilon \right)^2}+ \frac{\xi\left(1-2\beta\right)}{\left(x-\xi+i\epsilon \right)^2}+ \frac{4\xi\beta}{\left(x+\xi-i\epsilon \right)\left(x-\xi+i\epsilon \right)}\right)\,,\nonumber\\
    \mathcal{I}_T^{q_1} &=&\int_{-1}^{1} dx F^q\left(x,\xi,\Delta^2\right)\left( \frac{2\xi z (1-z)}{\left(x-\xi+i\epsilon \right)}- \frac{2\xi \beta (1-z)^2}{(1-\beta)\left(x+\xi-i\epsilon \right)} +\frac{2\xi (1-z)^2}{(1-\beta)\left(x-\xi\left(1-2\beta\right)-i\epsilon \right)} \right)\,,\nonumber\\
    \mathcal{I}_T^{q_2} &=&\int_{-1}^{1} dx F^q\left(x,\xi,\Delta^2\right)\left( \frac{2\xi \beta z^2}{(1-\beta)\left(x-\xi+i\epsilon \right)}- \frac{2\xi z(1-z)}{\left(x+\xi-i\epsilon \right)} -\frac{2\xi z^2}{(1-\beta)\left(x+\xi\left(1-2\beta\right)+i\epsilon \right)} \right)\,. \nonumber \\
    \label{IT}
\end{eqnarray}
Just as for the longitudinal photon case, these can be conveniently decomposed into Compton-type form factors. For the gluon case again we encounter second order poles, that require the $x$ derivative of the gluon GPD to be well behaved.
For the quark case, we observe that the relevant form factors will not be a function of $\xi$ and $\Delta^2$ alone, but also integrals having a dependence on the parameter $\beta$ appear.
We have shown all the relevant form factors and the expressions of integrals ${\cal I}^{q,g}_{L,T}$ in the Appendix \ref{sec:Integrals}.

Thus we are now equipped with all the integrals required in our $\gamma^{\ast}p$ subprocess amplitude in both longitudinal and transverse photon polarizations. In the left panel of Fig.\ref{fig:plot2} we see the gluon contribution to our process of interest and in the right panel we see the contribution from the quark exchanges.
\subsection{Cross section}
\label{sec:2c}
Our dijet cross section, summed over the parton helicities and color of the $q\bar{q}$ pair in the final state, can be written in the standard form, putting into evidence the contributions of transversely and longitudinally polarized photons and their relevant interferences:
\begin{eqnarray}
\frac{d\sigma (ep \to e'jj p)}{dy\, dQ^2 d \phi d\vec p_\perp^{~2} d \vec \Delta_\perp^2 dz} &=& \frac{\alpha_{em}}{\pi Q^2 y} \Bigg[\frac{1 + (1 - y)^2}{2}  \frac{d\sigma_T(\gamma^* p \to jj p)}{ d\vec p_\perp^{~2} d \vec \Delta_\perp^2 dz  }
+ (1-y) \frac{d\sigma_L(\gamma^* p \to jj p)}{ d\vec p_\perp^{~2} d \vec \Delta_\perp^2 dz  } \nonumber \\
&-& (2-y) \sqrt{1-y} \cos \phi \frac{d\sigma_{LT}(\gamma^* p \to jj p)}{ d\vec p_\perp^{~2} d \vec \Delta_\perp^2 dz  } \nonumber \\
&-& 2(1 - y) \cos 2\phi\, 
\frac{d\sigma_{TT}(\gamma^* p \to jj p)}{ d\vec p_\perp^{~2} d \vec \Delta_\perp^2 dz  } \Bigg],
\label{cs}
\end{eqnarray}
where $\phi$ is the azimuthal angle between the leptonic plane and the plane spanned by the outgoing dijets. 

For a given quark jet flavor $f$ we have for $i \in \{ \rm{T,L,LT,TT} \}$:
\begin{eqnarray}
 \frac{d\sigma_{i}(\gamma^* p \to jj p)}{ d\vec p_\perp^{~2} d \vec \Delta_\perp^2 dz  }  = \frac{\alpha_s^2 \alpha_{em} e_f^2}{16 \pi N_c (1- \xi^2)} \frac{\Omega_i}{(\vec p_\perp^{~2} + m_f^2 + z(1-z) Q^2)^4},
\end{eqnarray}
with
\begin{eqnarray}
    \Omega_T &=& m_f^2 |{\cal I}^g_L|^2 + \vec p_\perp^{~2} \Big( |(1-z) {\cal I}^g_T + 2 C_F {\cal I}^{q_1}_T|^2 + |z {\cal I}^g_T - 2 C_F {\cal I}^{q_2}_T|^2 \Big), \nonumber \\
   \Omega_L &=& 4 z^2 (1-z)^2 Q^2 \,   |{\cal I}_L^g + 2C_F {\cal I}_L^{q}|^2, 
\end{eqnarray}
and, for the interference pieces
\begin{eqnarray}  
   \Omega_{LT}  &=& 2 z (1-z) Q \, |\vec p_\perp|\, \, \Re e \left[ \left({\cal I}_L^g + 2C_F {\cal I}_L^{q}\right) \left( (1 - 2z) {\cal I}_T^g + 2C_F ({\cal I}_T^{q_1} + {\cal I}_T^{q_2}) \right)^{\ast} \right], \nonumber \\
   \Omega_{TT} &=& \vec p_\perp^{~2} \,  \Re e \left[ \left((1-z) {\cal I}_T^g + 2C_F {\cal I}_T^{q_1}\right)\left(z {\cal I}_T^g - 2C_F {\cal I}_T^{q_2}\right)^{\ast} \right].
   \label{cs2}
\end{eqnarray}

\subsection{Double distribution approach for GPD modeling}
\label{sec:2d}

For modeling GPDs, we use the DD representation, wherein the polynomiality of the resulting GPDs is warranted.
In the DD approach, the outgoing momenta flows in both the $t$ and $s$--channels (given by $\Delta^+$ and $p^+$) are treated independently with factors $\alpha$ and $\beta$ respectively which determine the fraction of momenta carried out by the quark pair and the parton. These variables in the context of DDs, represent the momentum asymmetry between initial and final states and the usual longitudinal momentum fraction.

To explicitly describe the distinct domains in the $\left(x,\xi\right)$ space for the integration over $\beta$, we can rewrite the above expression as

\begin{figure*}
    \centering   
    \includegraphics[width=0.45\linewidth]{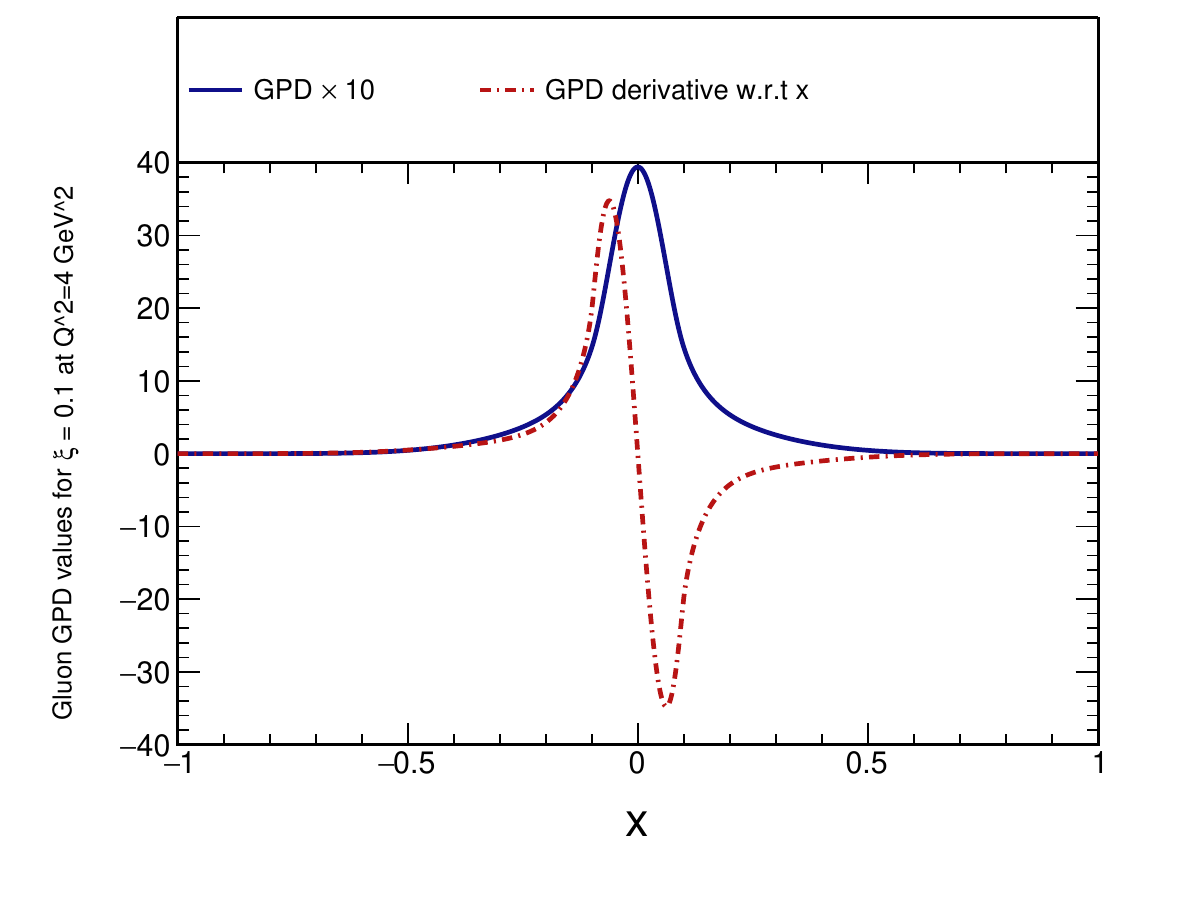} \hspace{1em}
    \includegraphics[width=0.45\linewidth]{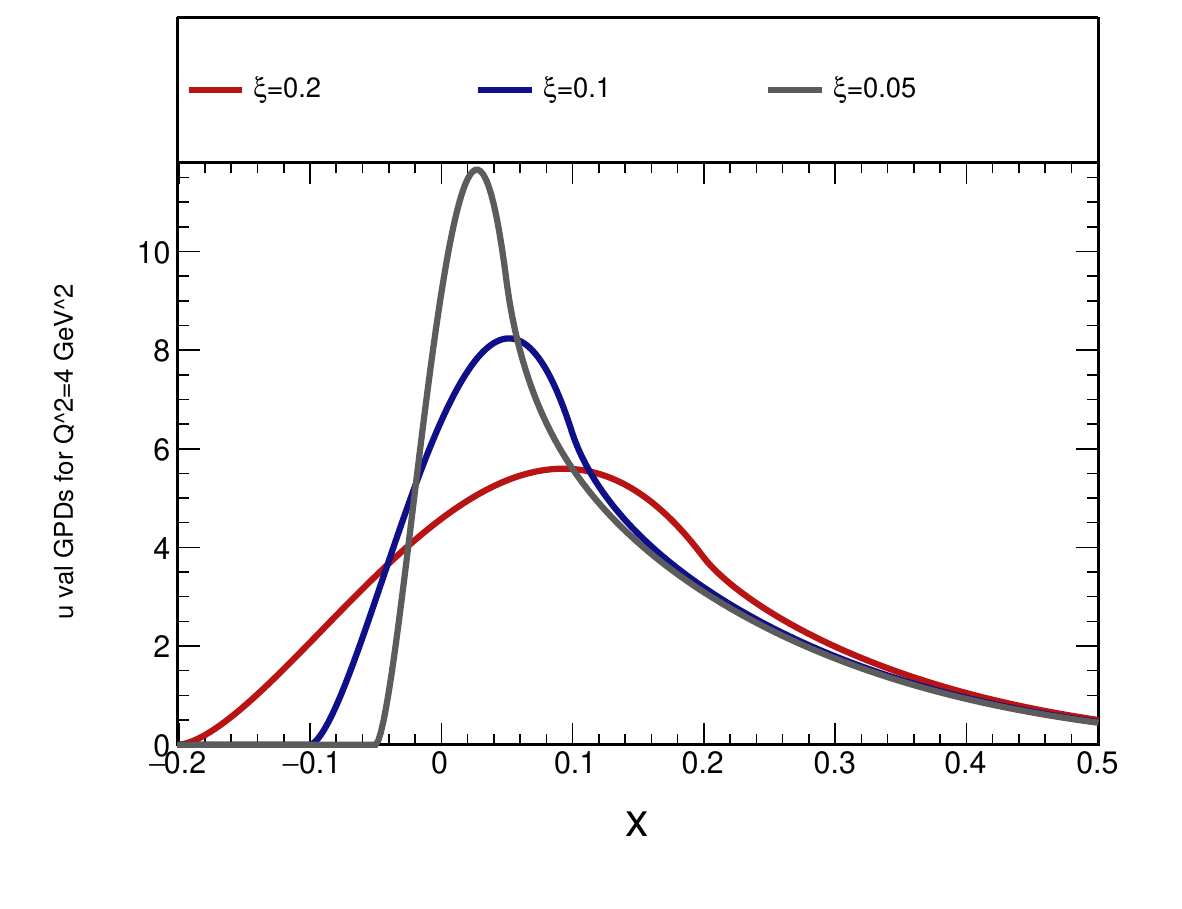} \hspace{1em}
    \includegraphics[width=0.45\linewidth]{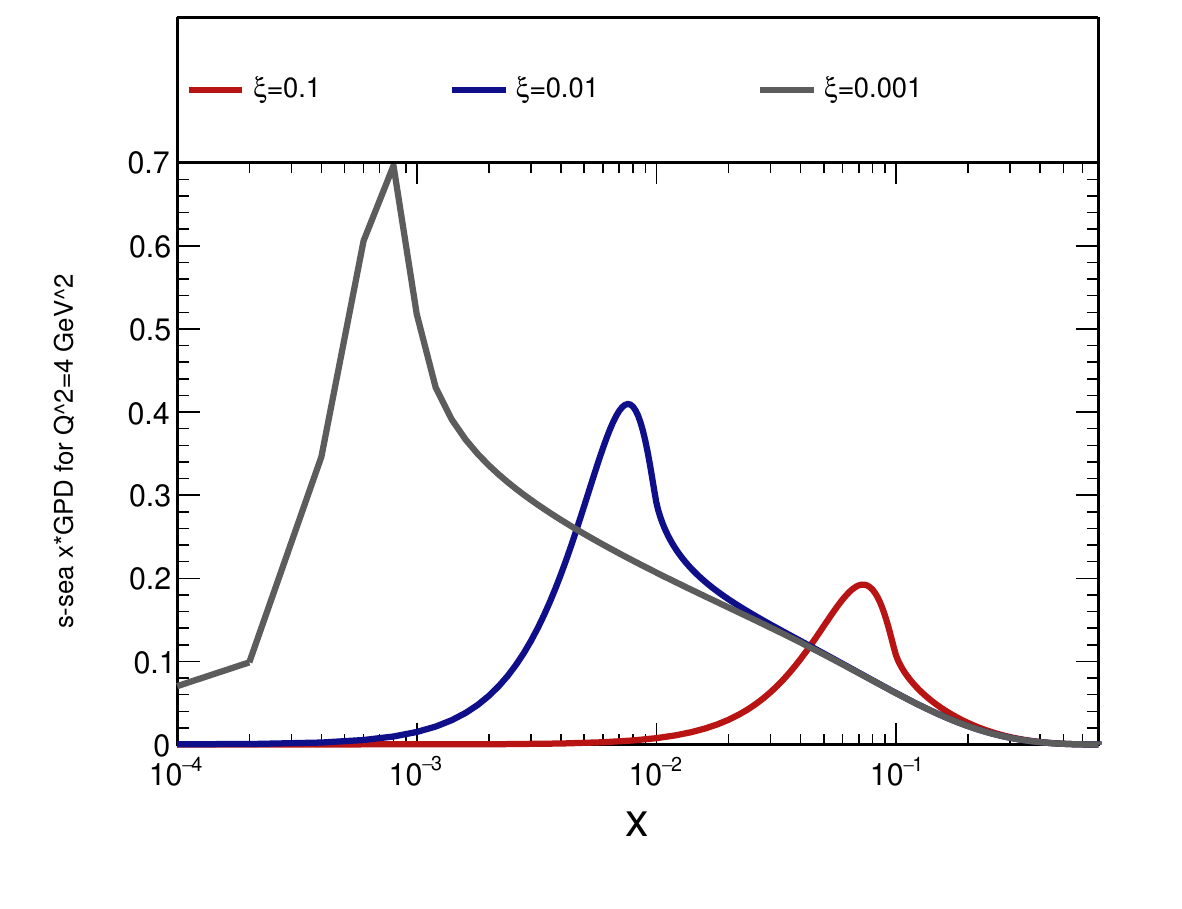}
    \caption{GPDs used in this work as a function of the momentum fraction $x$. The top-left panel shows the forward gluon GPD vs $x$ for $\xi=0.1$ as a solid blue line and its derivative with respect to $x$ as a dash-dotted red line, the top-right panel shows the forward u-valence quark GPDs vs $x$ at $\xi=0.2,0.1$ and $0.05$ with red, blue and black curves respectively, and the bottom panel shows the forward s-sea quark $x\cdot$GPD vs $x$ at $\xi=0.1,0.01$ and $0.001$ with red, blue and black curves, respectively.}
    \label{fig:plot3}
\end{figure*}

\begin{equation}
F^{i}\left(x,\xi,\Delta^2\right)=\int_{-1}^1 d\beta \frac{1}{\xi}\, g^{i}\left(\beta,\frac{x-\beta}{\xi},\Delta^2\right)\Xi\left(\beta|x,\xi\right)\,,
\end{equation}
where  $i \in \{{\rm gluon},{\rm sea},{\rm valence}\}$ and the function  $\Xi\left(\beta|x,\xi\right)$ divides the $\beta$ region into the so--called DGLAP and ERBL support region of the GPD as
\begin{eqnarray}
    \Xi\left(\beta|x,\xi\right)&=& \theta\left(x>\xi\right)\theta\left(\frac{x+\xi}{1+\xi}>\beta>\frac{x-\xi}{1-\xi}\right) + \theta\left(x<-\xi\right)\theta\left(\frac{x+\xi}{1-\xi}>\beta>\frac{x-\xi}{1+\xi}\right)\nonumber\\&+&\theta\left(-\xi<x<\xi\right)\theta\left(\frac{x+\xi}{1+\xi}>\beta>\frac{x-\xi}{1+\xi}\right)\,.
\end{eqnarray}
We have shown the gluon and quark GPDs used in this work as a function of momentum fraction $x$ in Fig.~\ref{fig:plot3}.
To model the DDs, we use the Radyushkin Ansätze with a factorized $\Delta^2$-dependence as \cite{Radyushkin:1998es, Radyushkin:1998bz, Goloskokov:2006hr},
\begin{equation}
g^i\left(\beta,\alpha,\Delta^2 \right)=  f_i\left(\beta\right)\pi_i\left(\beta,\alpha\right) \, {\cal F}_i (\Delta^2) \,,
\label{DDmodel}
\end{equation}
where
\[
f_i\left(\beta\right) = \left \{\begin{matrix}
     \left|\beta\right|y_i\left(\left|\beta\right|\right) \quad {\rm for\,\,i=gluon} \\ y_i\left(\left|\beta\right|\right)\sign \left(\beta\right) \quad {\rm for\,\,i=sea} \\ y_i\left(\beta\right)\theta \left(\beta\right) \quad {\rm for\,\,i=valence}.
\end{matrix}\right.
\]
Here $y_i$ are the usual PDFs corresponding to the $i$-th flavor of the partons exchanged for the dijet production.

The profile function $\pi_i\left(\beta,\alpha\right)$ in Eq.~\eqref{DDmodel} introduces the skewness dependence via $\alpha$ and can be chosen as
\begin{equation}
\pi_i\left(\beta,\alpha\right)=\frac{\Gamma(2n_i + 2)}{2^{2n_i+1} \, \Gamma^2(n_i + 1)} 
\frac{[(1 - |\beta|)^2 - \alpha^2]^{n_i}}{(1 - |\beta|)^{2n_i + 1}},
\end{equation}
where
\[
n_i = \left \{\begin{matrix}
     1 \quad {\rm for\,\,i=valence} \\ 2\quad {\rm for\,\,i=gluon,\,sea}
\end{matrix}\right.
\]
such that the profiles are normalized $\int_{-1+\left|\beta\right|}^{1-\left|\beta\right|}d\alpha\,\pi_i\left(\beta,\alpha\right)=1$ and the quark and gluon profiles are antisymmetric and symmetric in $\alpha$ respectively.
Similar profiles were used in the successful GPD phenomenology of \citet{Goloskokov:2006hr,Goloskokov:2007nt,Goloskokov:2009ia} for a number of reactions.
As far as the collinear parton input is concerned, we use the CT18 next-to-leading order (NLO) PDF set from \cite{Hou:2019efy}.
For the $\Delta$--dependence, we use for sea quarks and gluons a parametrization
${\cal F}(\Delta^2) = \exp(B t/2)$ (recall that $\Delta^2 = t$ is negative), with a common slope parameter $B = 4 ~\rm{GeV}^{-2}$.
For valence quarks, we follow \cite{Belitsky:2001ns}, by using 
\begin{eqnarray}
  {\cal F}_{\rm val}^u(\Delta^2) = F_1^{p}(\Delta^2) + \frac{1}{2}F_1^{n}(\Delta^2) , \quad    {\cal F}_{\rm val}^d(\Delta^2) = F_1^{p}(\Delta^2) +  2 F_1^{n}(\Delta^2) . 
\end{eqnarray}
Here $F_1^{p,n}$ are the Dirac form factors of the proton and neutron, respectively, which can be expressed in terms of the Sachs electric and magnetic form factors as
\begin{eqnarray}
    F_1^{p,n}\left(\Delta^2\right)=\left(1+\frac{|\Delta^2|}{4m_{p,n}^2}\right)^{-1}\left(G_E^{p,n}\left(\Delta^2\right)+\frac{|\Delta^2|}{4m_{p,n}^2}G_M^{p,n}\left(\Delta^2\right)\right)\,,
\end{eqnarray}
where $m_{p,n}$ denotes the proton and neutron mass, and $G_{E,M}^{p,n}$ are the Sachs electric and magnetic form factors for the proton and the neutron given by
\begin{eqnarray}
    G_{E}^{p}\left(\Delta^2\right)=\left(1+\frac{|\Delta^2|}{\Lambda^2}\right)^{-1}\,,\quad G_{M}^{p,n}\left(\Delta^2\right)=\mu_{p,n}G_{E}^{p}\left(\Delta^2\right)\,,\quad G_{E}^{n}\left(\Delta^2\right)=0\,,
\end{eqnarray}
where $\Lambda^2=0.71\,\rm GeV^2$, $\mu_p=2.793$ and $\mu_n=1.913$.
The so-constructed quark GPDs fulfill the important sum rule
\begin{eqnarray}
\sum_q e_q \int_{-1}^1 dx \, H^q(x,\xi,\Delta^2) = F_1^{p}(\Delta^2) \, .    
\end{eqnarray}

\section{Results and discussions}
\label{sec:3}
\subsection{Results for an extensive phase space}
\label{sec:3a}
\begin{figure*}[t]
    \centering   
    \includegraphics[width=0.48\linewidth]{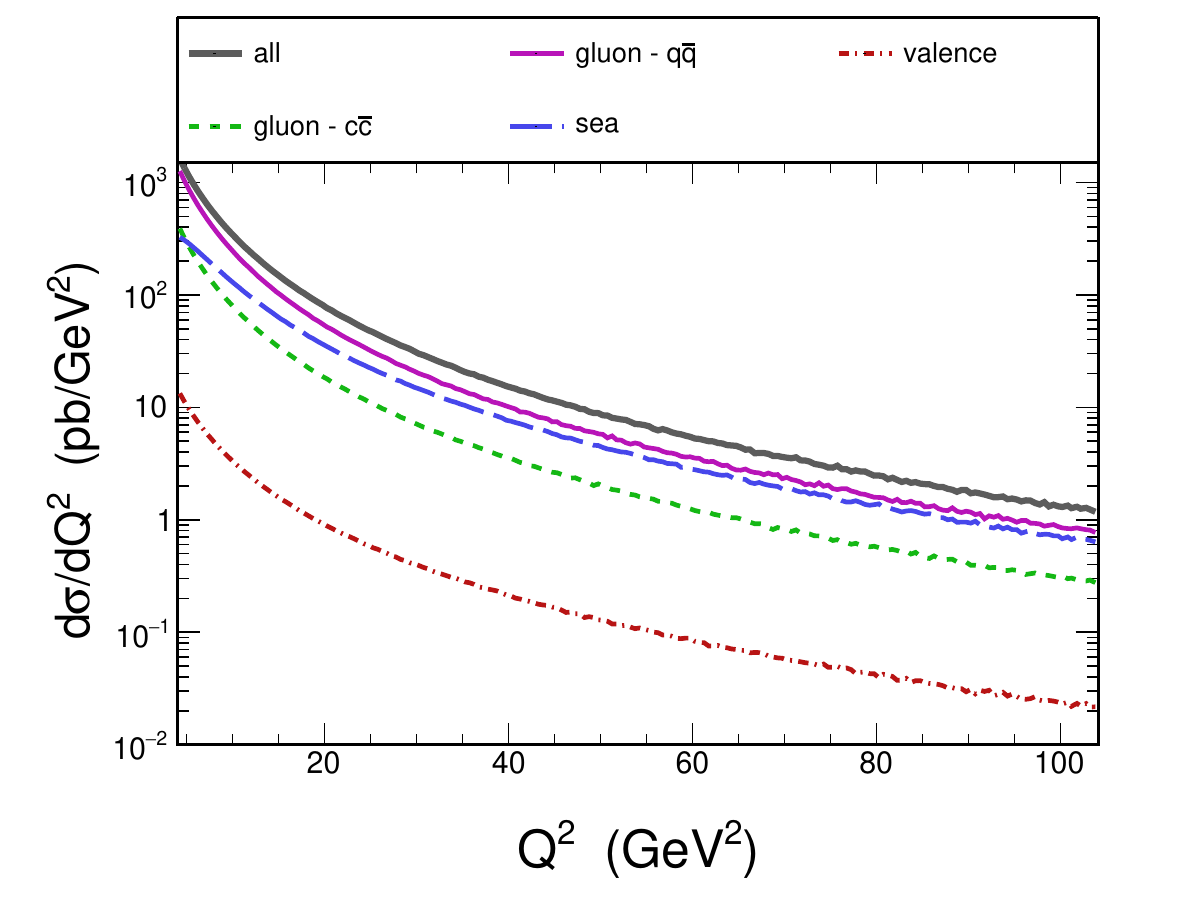} \hspace{1em}
    \includegraphics[width=0.48\linewidth]{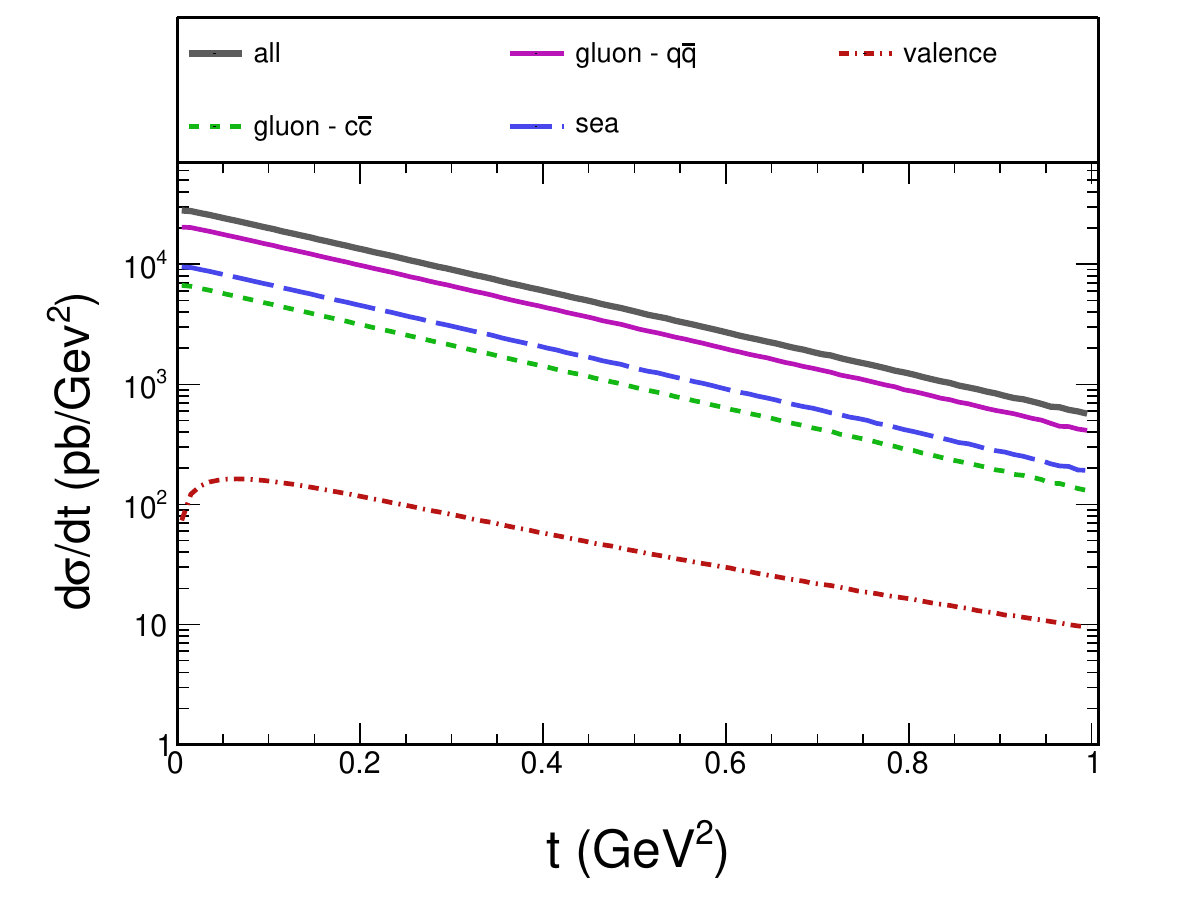}
    \caption{The left panel shows the $Q^2$ distribution of the cross section whereas the right panel shows the distribution in the Mandelstam variable $t$. The solid black curve represents the total contribution from sea and valence light $u,d,s$ quark exchanges as well as gluon exchanges, including both light quark-antiquark production and the heavy $c\bar{c}$ final state. The solid magenta curve shows the gluon contribution, the red dash-dotted curve corresponds to the valence quark contribution, the blue long-dashed curve to the sea quark contribution, and the green short-dashed curve to the gluon-induced contribution to the $c\bar{c}$ final state.}
    \label{fig:Qt}
\end{figure*}
\begin{figure*}[t]
    \centering   
    \includegraphics[width=0.48\linewidth]{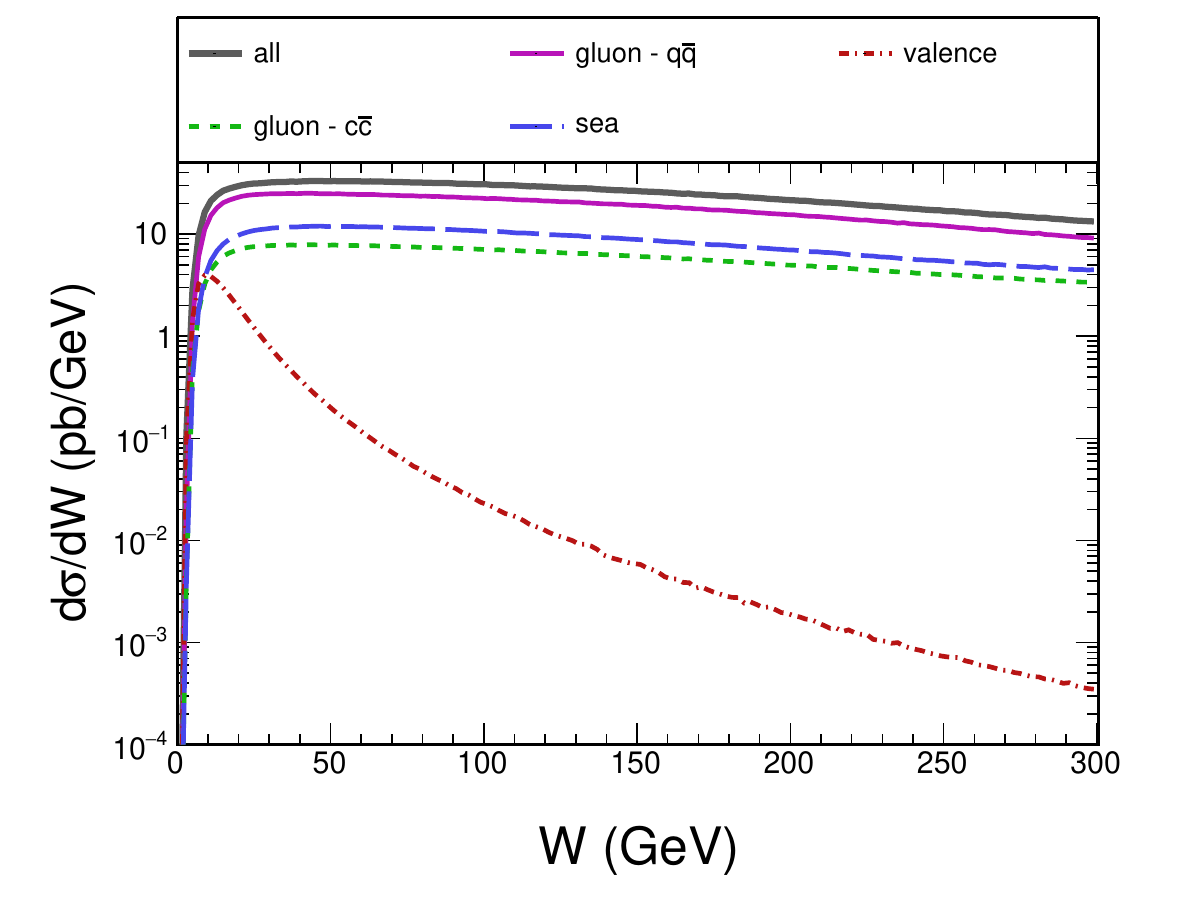} \hspace{1em}
    \includegraphics[width=0.48\linewidth]{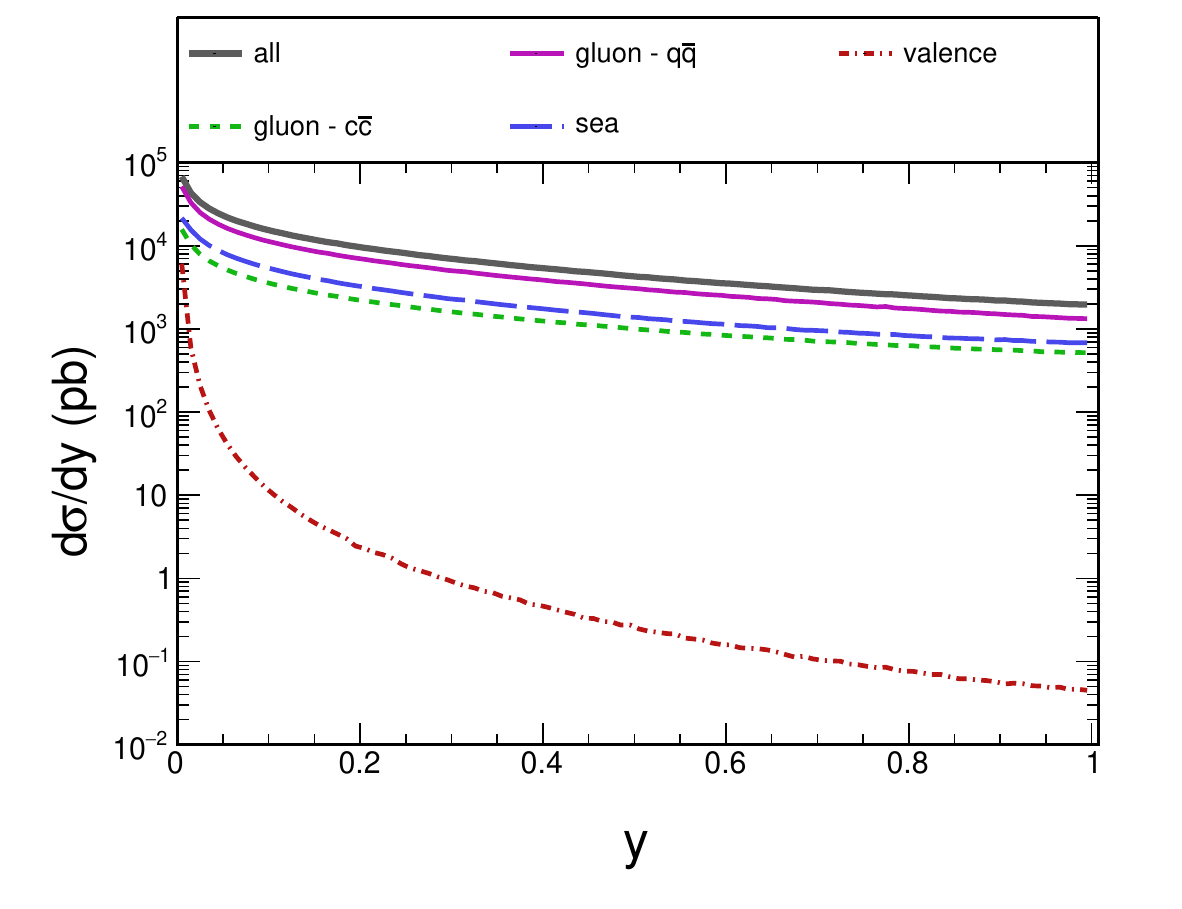}
    \caption{The left panel shows the dependence of the cross section on the invariant center of mass energy of the $\gamma^{\ast}p$ subprocess $W$ whereas the right panel shows the distribution in inelasticity variable $y$. Identification of the curves is the same as in Fig.~\ref{fig:Qt}.}
    \label{fig:wy}
\end{figure*}
\begin{figure*}[t]
    \centering   
    \includegraphics[width=0.48\linewidth]{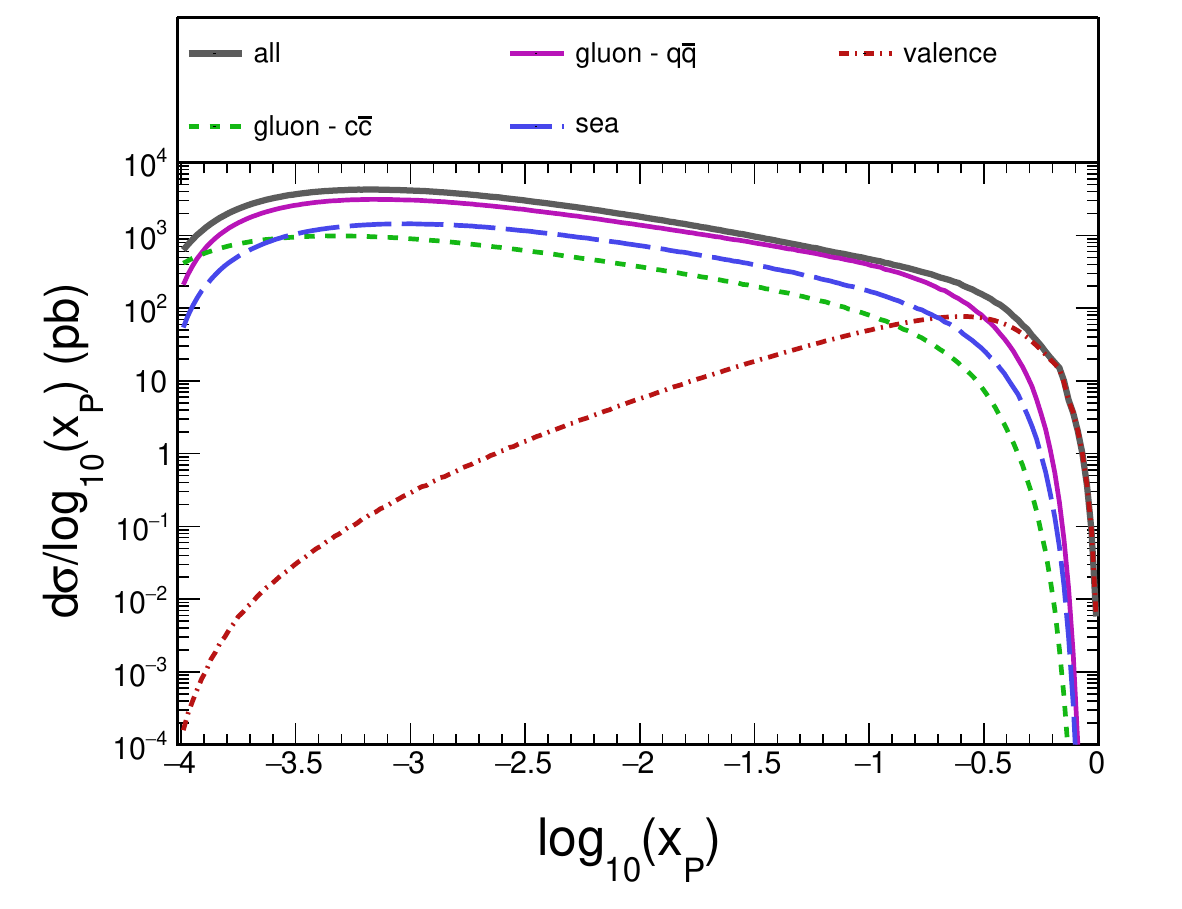} \hspace{1em}
    \includegraphics[width=0.48\linewidth]{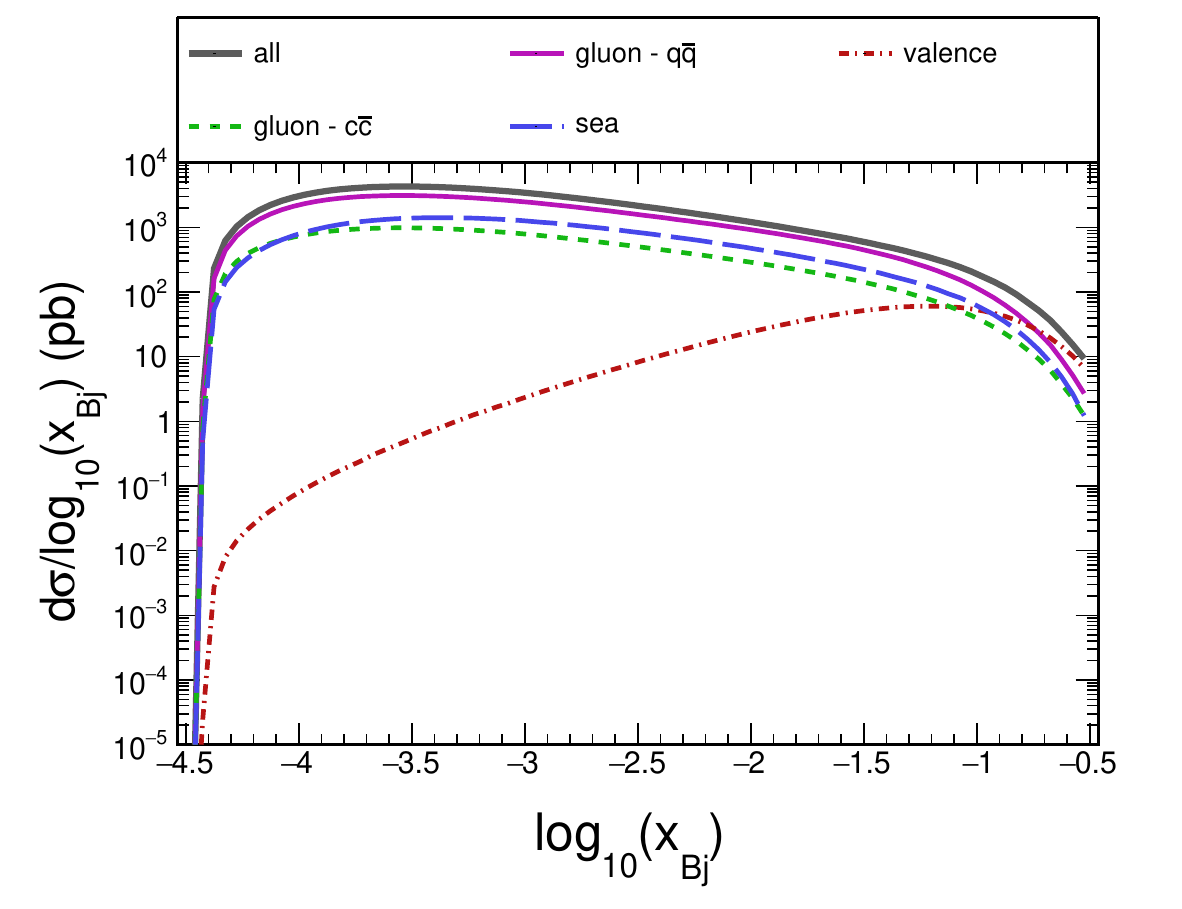}
    \caption{The left panel shows the $x_{\mathbb{P}}$ distribution of the cross section whereas the right panel shows the distribution in the Bjorken variable $x_{\rm Bj}$. Identification of the curves is the same as in Fig.~\ref{fig:Qt}.}
    \label{fig:x}
\end{figure*}
\begin{figure*}[t]
    \centering   
    \includegraphics[width=0.48\linewidth]{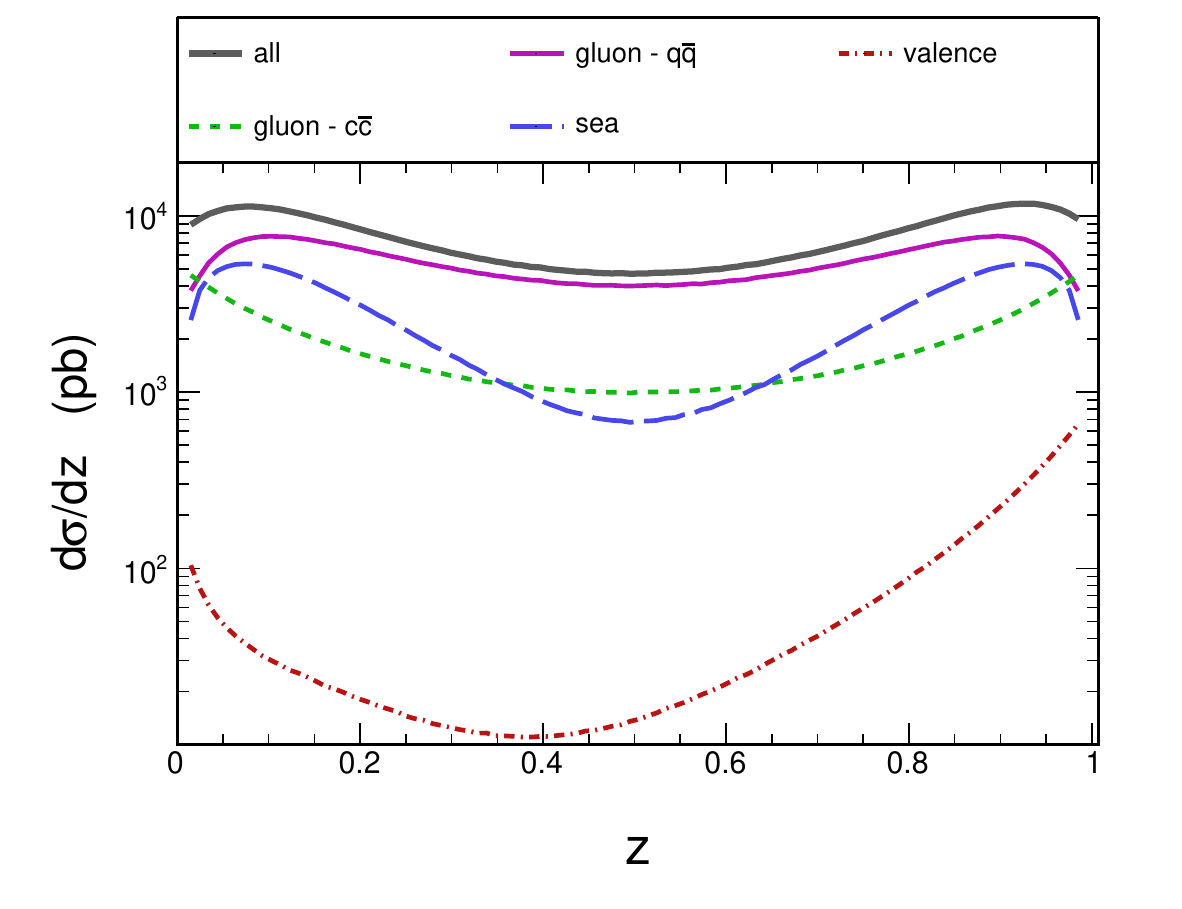}\hspace{1em}    \includegraphics[width=0.48\linewidth]{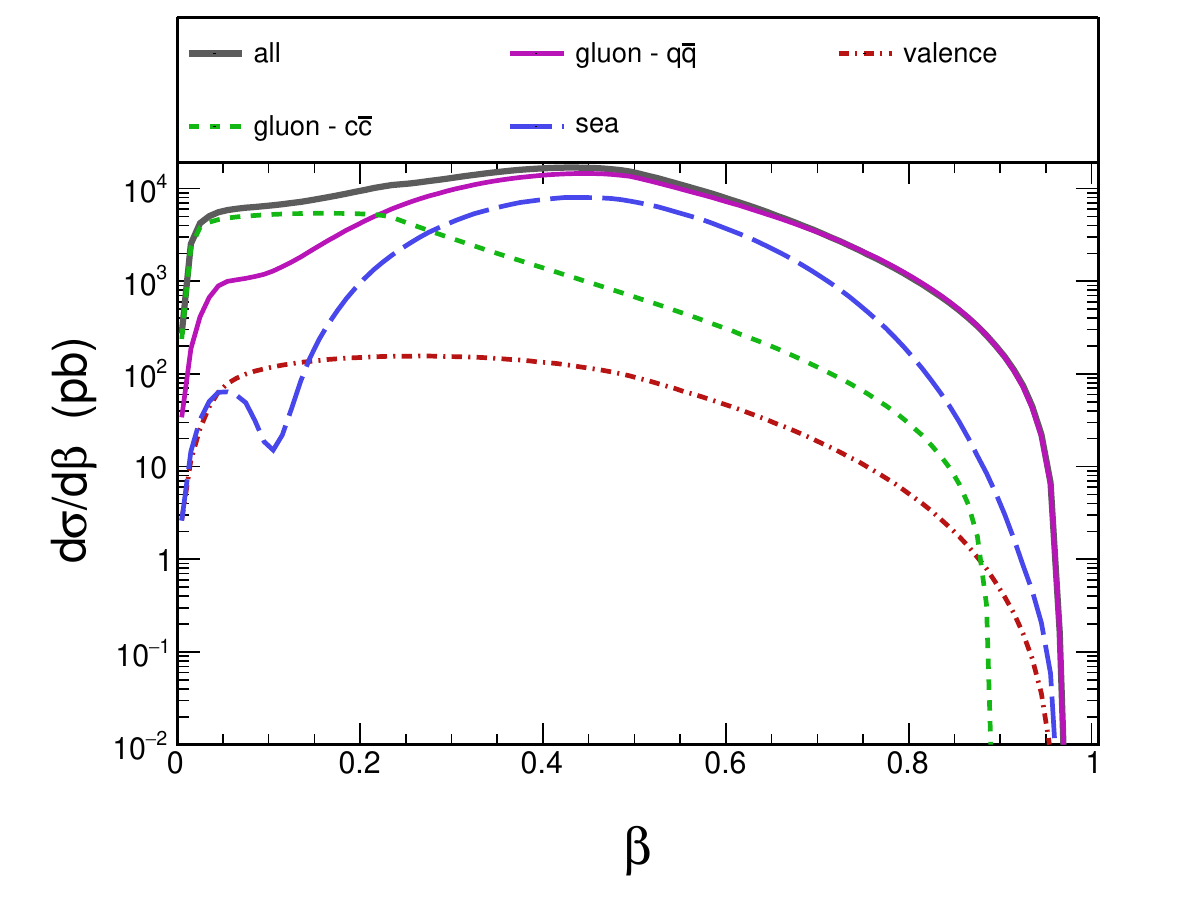}   
    \caption{The left panel shows the cross section distribution in the longitudinal momentum fraction $z$ whereas the right panel shows the distribution in the  $\beta$ parameter. Identification of the curves is the same as in Fig.~\ref{fig:Qt}.}
    \label{fig:zbeta}
\end{figure*}

In this section, we will present the cross section for our process 
$ep \to e' q \bar q p$
for center of mass energy $\sqrt{s}=319$ GeV (HERA energy scale) in a broad range of kinematical variables, without any imposed experimental cuts, so that we are not solely confining ourselves to the diffractive kinematic regime.
The aim of this work is to identify the kinematic regions of interest where different contributions across gluons, sea and valence light $u,d,s$ quarks and their heavy counterparts such as charm  may play a crucial role.
To explore an extensive phase space we take $4<Q^2<104\rm\, GeV^2$, $0<y<1$, $0.01<z<0.99$, $1<p_{\perp}<15\,\rm GeV$ and $0<\Delta_{\perp}<3\,\rm GeV$.
In the calculation of the cross section as differential distributions presented in this work, we have used the CT18 NLO PDF set \cite{Hou:2019efy} at a factorization scale $\mu^2 =4 \, \rm{GeV}^2$ as input in the DD approach for modeling GPDs.

In Fig.~\ref{fig:Qt} we show the distributions in photon virtuality $Q^2$ and four-momentum transfer squared $t$ given in Eq.~\eqref{tmand}. 
Here the rest of the phase space variables are integrated over the aforementioned kinematical ranges.
In both cases the shapes of distributions are similar for different components (light sea and valence quarks, and gluons for light $q\bar{q}$ and heavy $c\bar{c}$). The contribution from the valence quarks ($u_{\rm val}, d_{\rm val}$) is small. We further investigate whether these components may be visible in some corner of the phase space.
In Fig.~\ref{fig:wy} we show distributions in center of mass energy $W$ in the $\gamma^{\ast} p$ system and inelasticity $y$. In both cases, the valence contribution shows a very different behavior from the gluon and sea quark exchanges. It does play some role at small $W$, then drops toward large $W$, as expected from a secondary Regge exchange.
Correspondingly, the contribution of valence quarks plays a role only at small values of the inelasticity $y$.
\begin{figure*}[t]
    \centering   
    \includegraphics[width=0.48\linewidth]{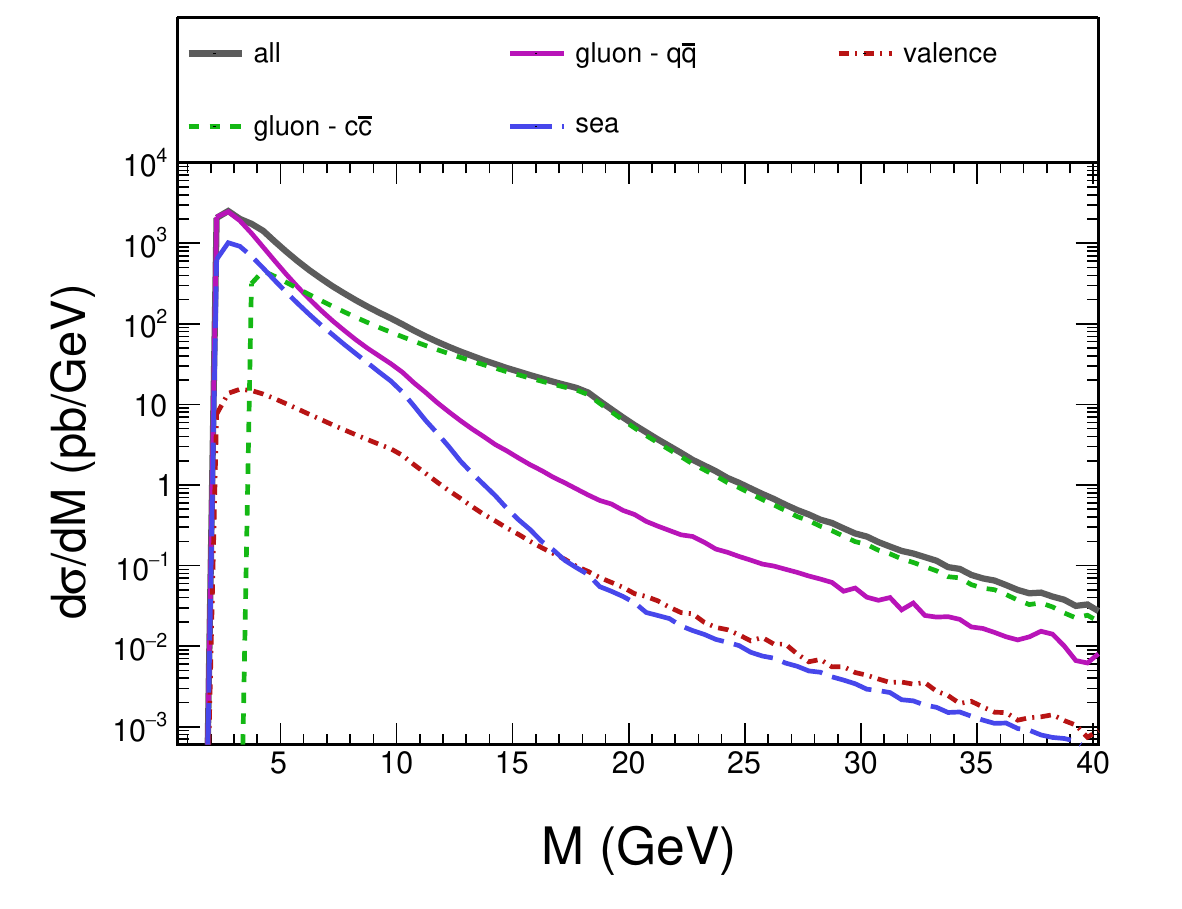} \hspace{1em}
    \includegraphics[width=0.48\linewidth]{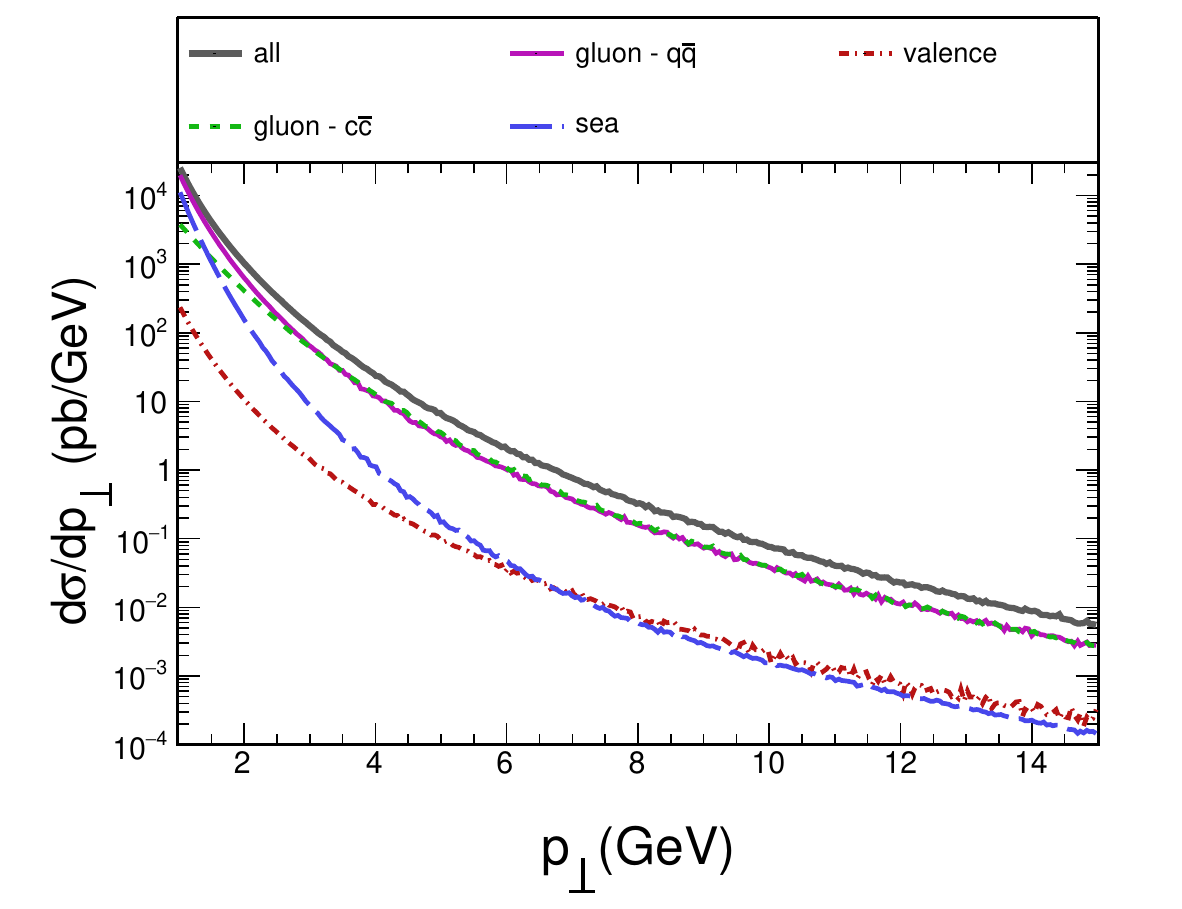}
    \caption{The left panel shows the dependence of the cross section on the invariant mass of the dijets $M$ whereas the right panel shows the distribution in the jet transverse momentum $p_{\perp}$. Identification of the curves is the same as in Fig.~\ref{fig:Qt}.}
    \label{fig:mpt}
\end{figure*}
\begin{figure*}[t]
    \centering   
    \includegraphics[width=0.65\linewidth]{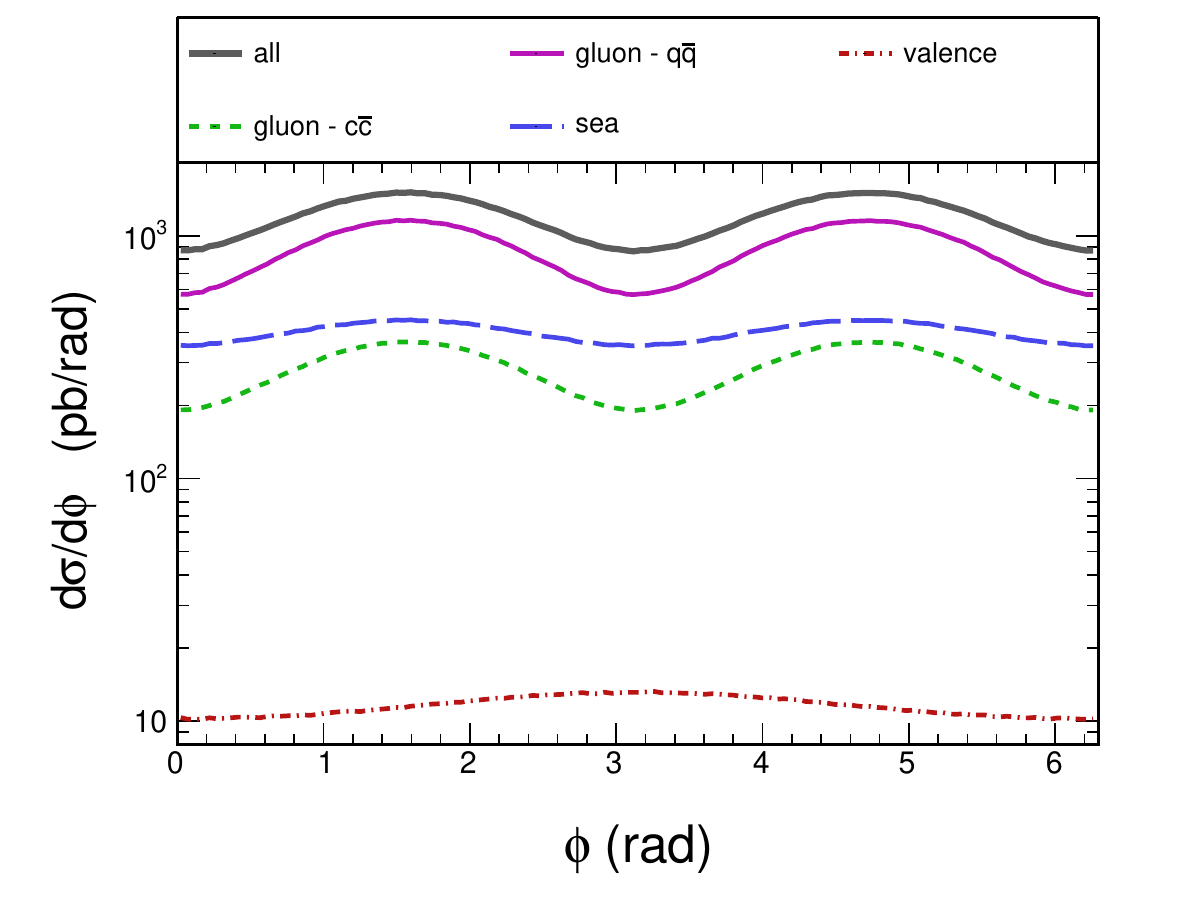}
    \caption{Cross section distribution in the azimuthal angle $\phi$ without any experimental cuts imposed. Identification of the curves is the same as in Fig.~\ref{fig:Qt}.}
    \label{fig:phim}
 \end{figure*}
\begin{figure*}[t]
    \centering   
    \includegraphics[width=0.65\linewidth]{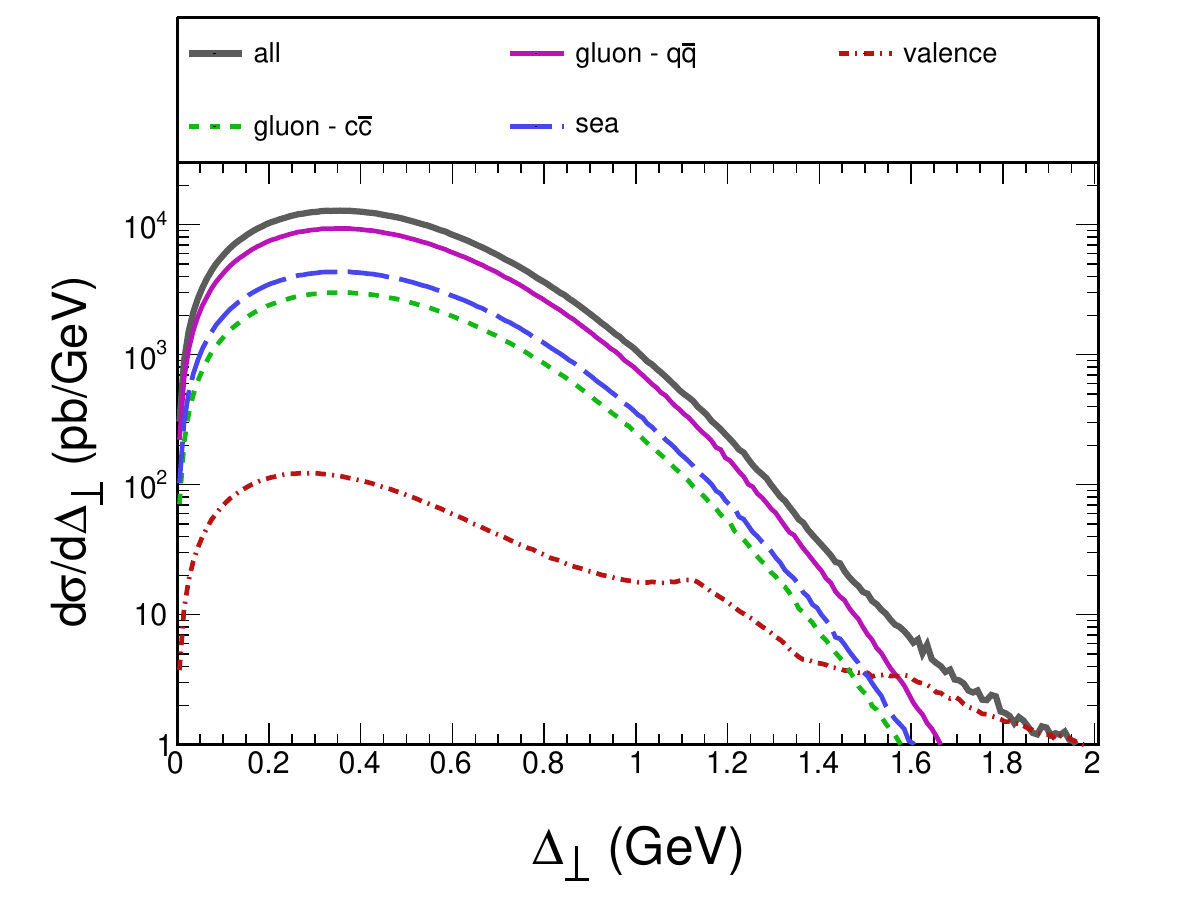}
    \caption{The figure shows the distribution in jet transverse momenta $\Delta_{\perp}$. Identification of the curves is the same as in Fig.~\ref{fig:Qt}.}
    \label{fig:del}
\end{figure*}
\begin{figure*}[h]
    \centering   
    \includegraphics[width=0.45\linewidth]{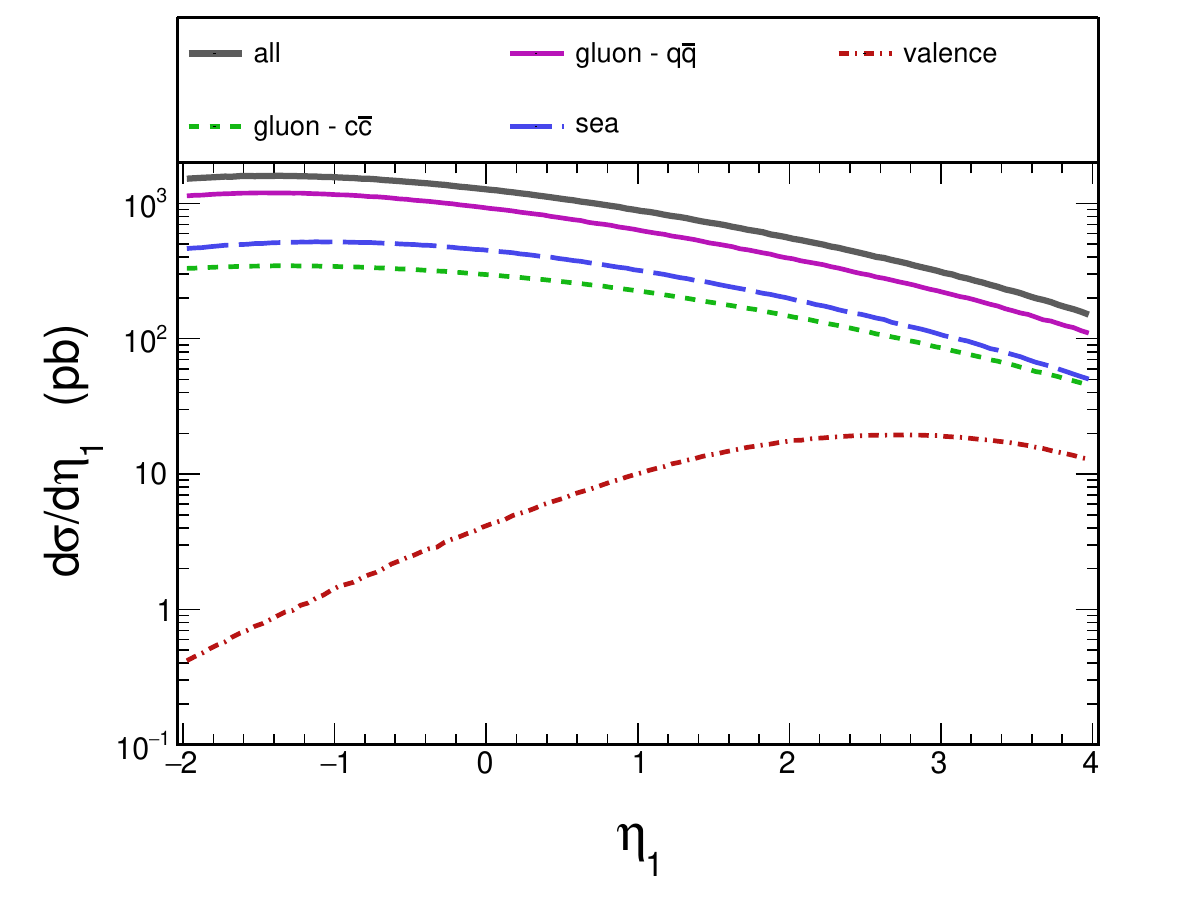} \hspace{1em}
    \includegraphics[width=0.45\linewidth]{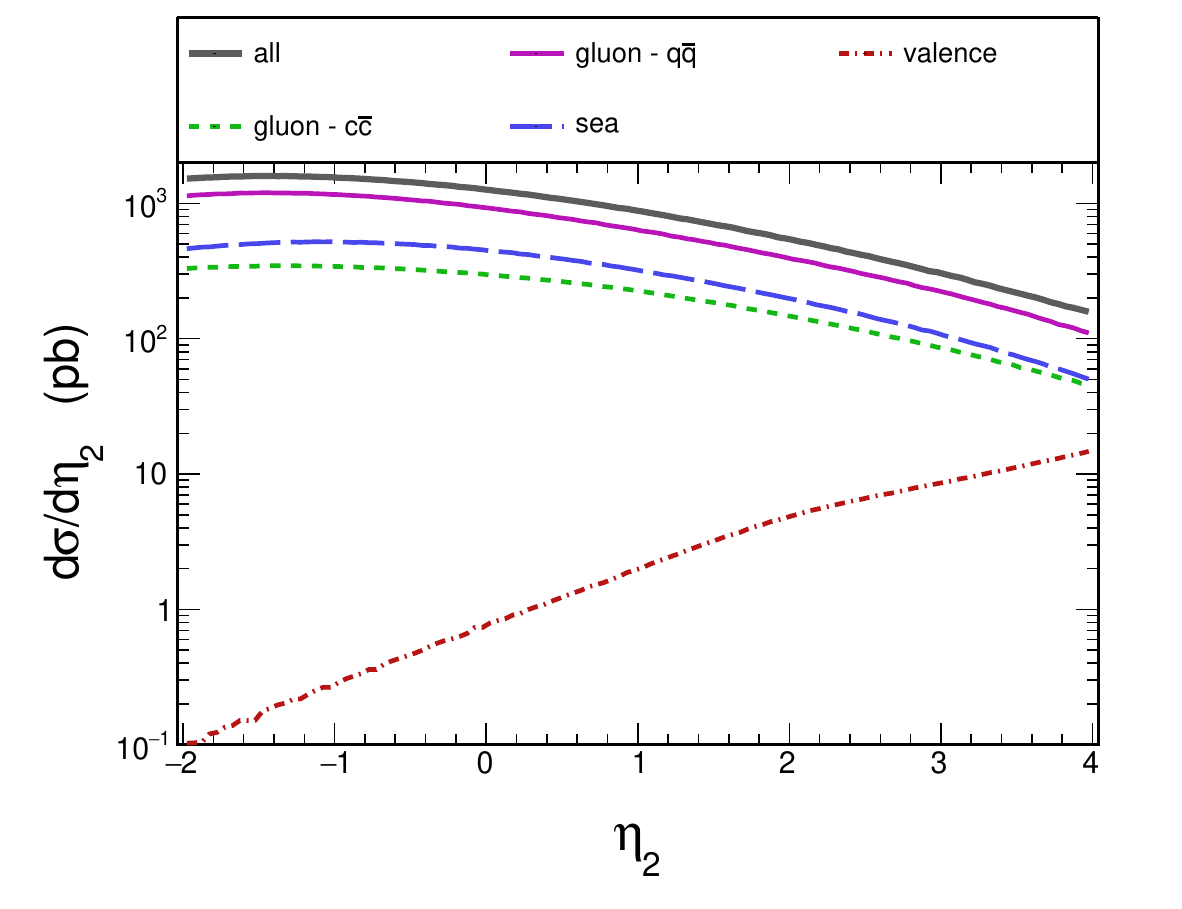} \hspace{1em}
    \includegraphics[width=0.45\linewidth]{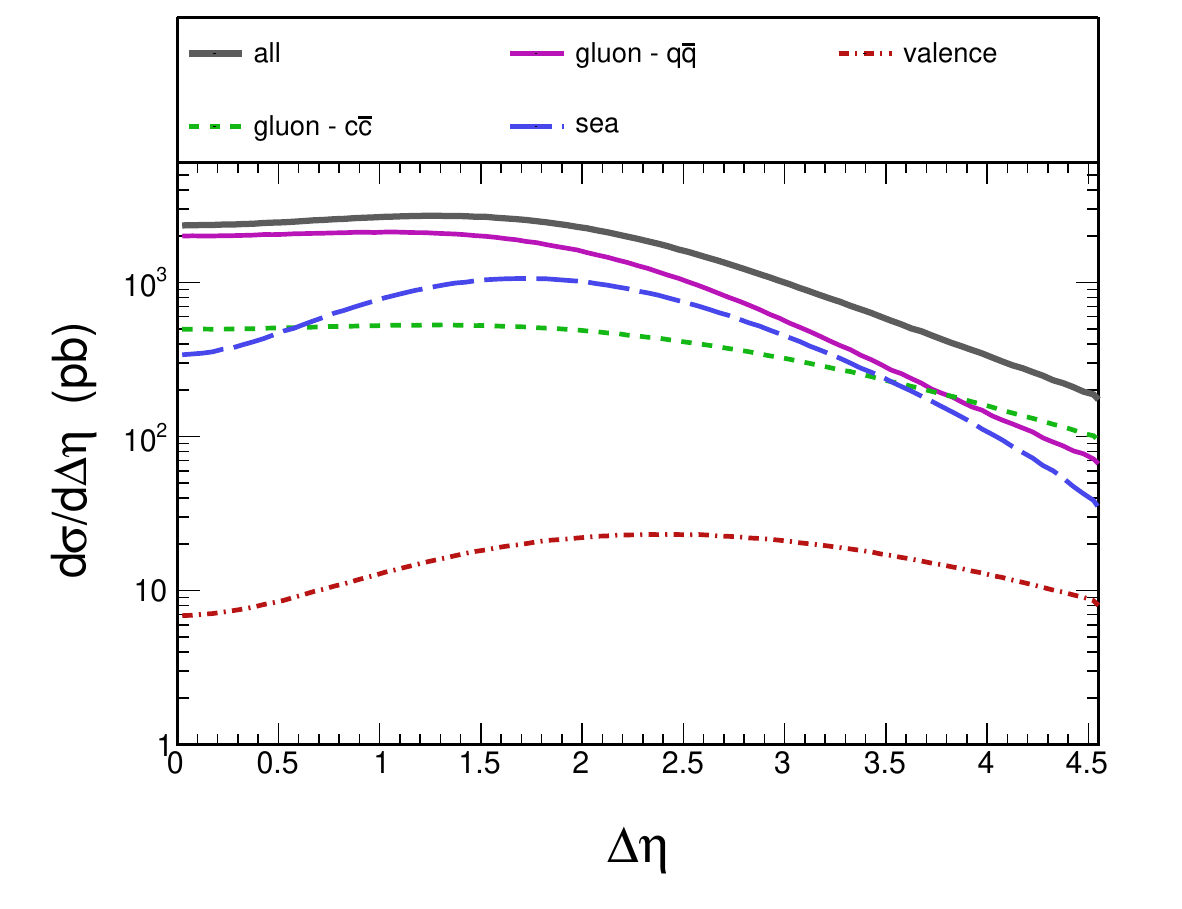}
    \caption{The top-left panel and the top-right panel shows the dependence of the cross section on the jet rapidities $\eta_1$ and $\eta_2$ respectively and the bottom panel shows the distribution in their difference $\Delta\eta$. Identification of the curves is the same as in Fig.~\ref{fig:Qt}.}
    \label{fig:eta}
\end{figure*}
\begin{figure*}[h]
    \centering   
    \includegraphics[width=0.99\linewidth]{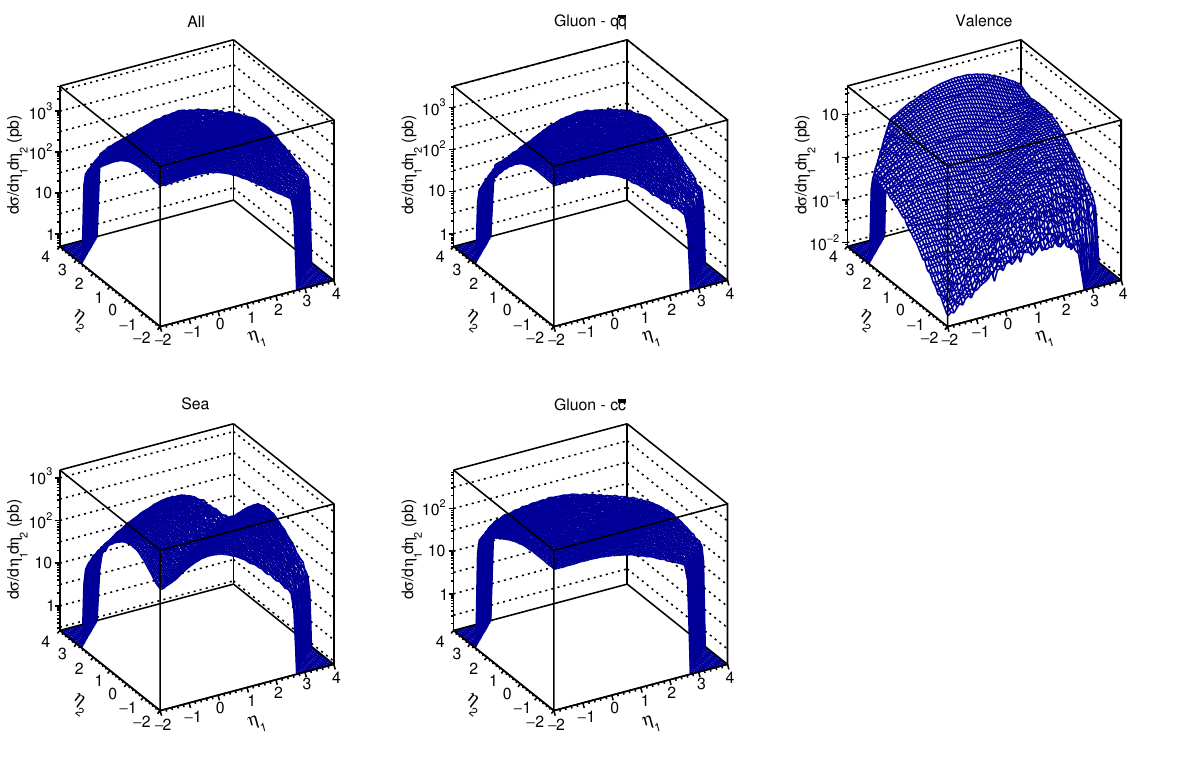} 
    \caption{The above figure shows two-dimensional distributions in $\eta_1$ and $\eta_2$. The top-left panel shows the total contribution from sea and valence light $u,d,s$ quark exchanges as well as gluon exchanges, including both light $q\bar{q}$ production and the heavy $c\bar{c}$ final state, the top-middle panel shows the contribution from gluon exhanges for light $q\bar{q}$ production, the top-right panel shows the valence quark contribution, the bottom-left panel shows the sea quark contribution and the bottom-middle panel shows the contribution from charm quarks.}
    \label{fig:eta2}
\end{figure*}
In Fig.~\ref{fig:x} we show the distribution in the fraction
of momentum of the proton carried by the color-singlet exchange, $x_{\mathbb{P}}$.
At the HERA experiments the main interest was concentrated on small $x_{\mathbb{P}}$,
i.e. on the so-called diffractive processes with large rapidity gap. In that case, one usually imposes a cut like $x_{\mathbb{P}}<0.1$. Here, we intentionally extended our calculation to larger $x_{\mathbb{P}}$ wherein we see that the valence quark contributions start dominating for $x_{\mathbb P} \gtrsim 0.3$. For completeness, in the right panel of Fig.6 we show the distribution in the Bjorken variable $x_{\rm Bj}$ which is qualitatively similar to that for $x_{\mathbb{P}}$.

In Fig.~\ref{fig:zbeta} we show the cross section distribution in the quark momentum fraction $z$ in the left panel.
Here, gluon and sea quark contributions are symmetric under $z \leftrightarrow 1-z$.
The contribution of valence quarks, however shows a strong asymmetry around $z = 0.5$, which translates to a forward-backward asymmetry in the quark polar angle in the dijet center of mass frame. The origin of this asymmetry is the interference of $t$--channel exchanges of positive and negative $C$--parity. As overall the valence contribution is small, the asymmetry is invisible in the full cross section.

Experimentally, one is not able to distinguish quark and antiquark jets separately, therefore the experimentally obtained jet momentum fraction distributions are symmetric. When comparing to such experimentally obtained data, one would need to symmetrize the $z$-distributions shown in the left panel of Fig.~\ref{fig:zbeta}.
Presently, such experimental data in jet momentum fraction distribution for the exclusive production of dijets are not available.

The distributions in $\beta$ shown in the right panel of Fig.~\ref{fig:zbeta} are somewhat irregular, and it would be good to measure them experimentally. This will be possible at Electron-Ion Collider (EIC) and perhaps at FCC$_{eh}$ in the future. 
Here we show the results calculated for the full phase space, and in the next section we also show the $\beta$ distribution with cuts that correspond to the ZEUS kinematical region so that we can tally our results with the available ZEUS data. An important note here is that for the results of the $c\bar{c}$ final state through gluon exchanges, we factor in the charm mass in the definition of the diffractive DIS $\beta$ parameter as in Eq.~\eqref{beta}.

Fig.~\ref{fig:mpt} shows the distribution in dijet invariant mass in the left panel and jet transverse momentum (identical for both jets in collinear approximation) in the right panel.
The sea  quark contribution becomes comparable to the gluon contribution at small $M$ and small $p_{\perp}$.

The azimuthal correlations between the leptonic and hadronic planes shown in Fig.~\ref{fig:phim} are particularly interesting. The interference terms shown in Eq.~\eqref{cs2} lead to sizable azimuthal-angle modulations, as seen in Fig.~\ref{fig:phim}. The biggest modulations are predicted for the gluon and sea components.
The small valence contribution has quite a different pattern from all other ones.

In Fig.~\ref{fig:del} the differential distribution in transverse momentum of the dijet system is shown.
We can see here that in the region $\Delta_{\perp}>1.6\,\rm GeV$, only the contribution from valence quark exchanges is predominant.

Finally, we present rapidity distributions in the jet rapidities $\eta_1$ and $\eta_2$ in Fig.~\ref{fig:eta}, top-left and top-right panels, and the distribution in rapidity difference $\Delta\eta$ in the bottom panel.
In Fig.~\ref{fig:eta2} we present two-dimensional distributions in the rapidities of both jets. We show separate distributions for each component separately. We can clearly notice that only in the valence case (shown in the top right panel), the contribution is not symmetric with respect to $\eta_1 \leftrightarrow  \eta_2$, the reason for the asymmetry being the same as for the $z$ distribution. Here the same caveat as for the momentum fraction distributions applies, when one wants to compare to jet rapidity distributions.

\subsection{Results for ZEUS kinematics}
\label{sec:3b}
We have also calculated the cross section distributions in the kinematical regions corresponding to both the H1 and ZEUS experiments, but we found that there is almost no region of the phase space where our distributions are in agreement with the available H1 data, confirming the findings of \cite{Linek:2024dzs}. This is understandable, as H1 data do not entail the exclusivity of the dijet production. So, we particularly focus on ZEUS kinematics.
In this section, we show the differential cross sections in the kinematic region where ZEUS data are available, applying experimental cuts corresponding to the ZEUS ($\sqrt{s}=318$ GeV) kinematical region of \cite{ZEUS:2015sns}  $Q^2>25$ GeV$^2$, $90<W<250$ GeV, $x_{\mathbb{P}}<0.01$, $M>5$ GeV, $p_{\perp}>2$ GeV, $0.1<y<0.64$ and, $\eta_{1,2}<2$.
\begin{figure*}[!htb]
    \centering   
    \includegraphics[width=0.45\linewidth]{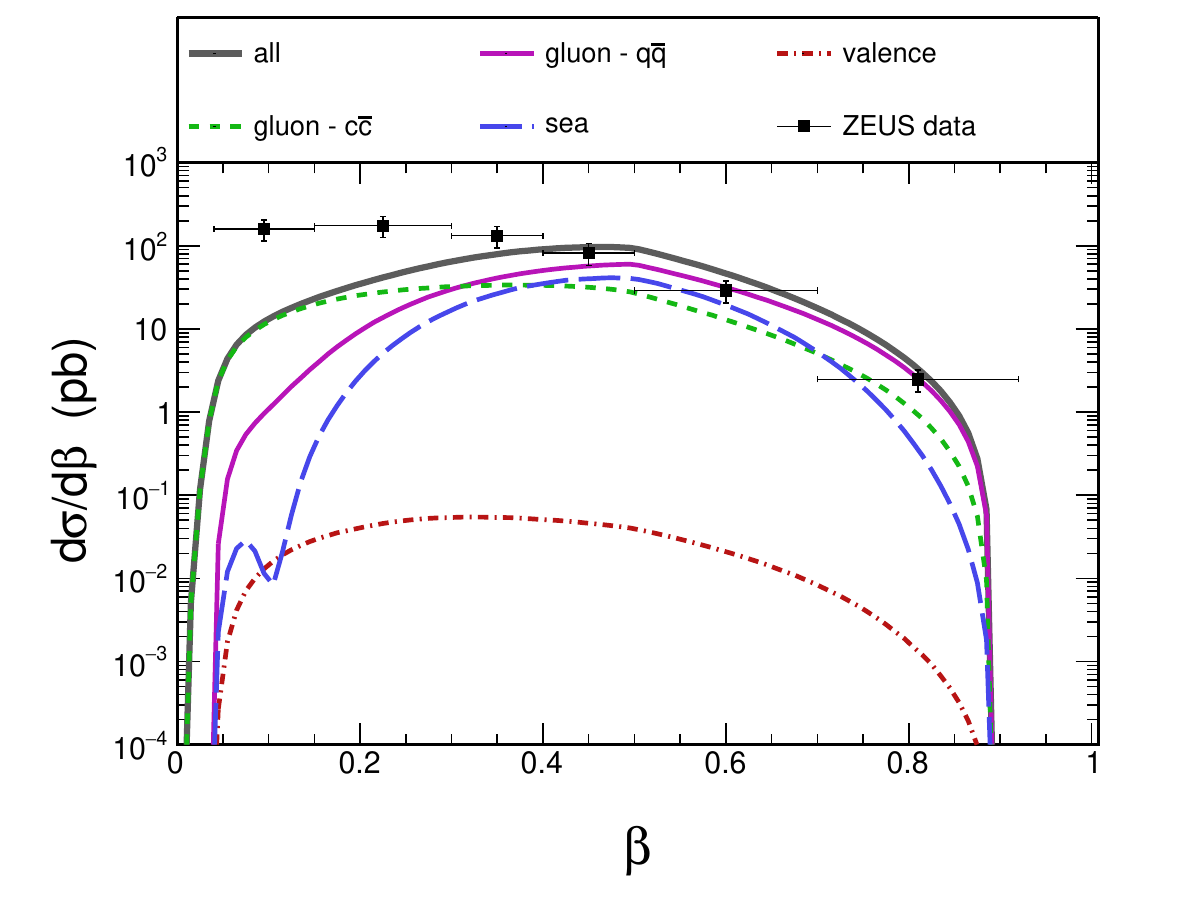} \hspace{1em} 
    \includegraphics[width=0.45\linewidth]{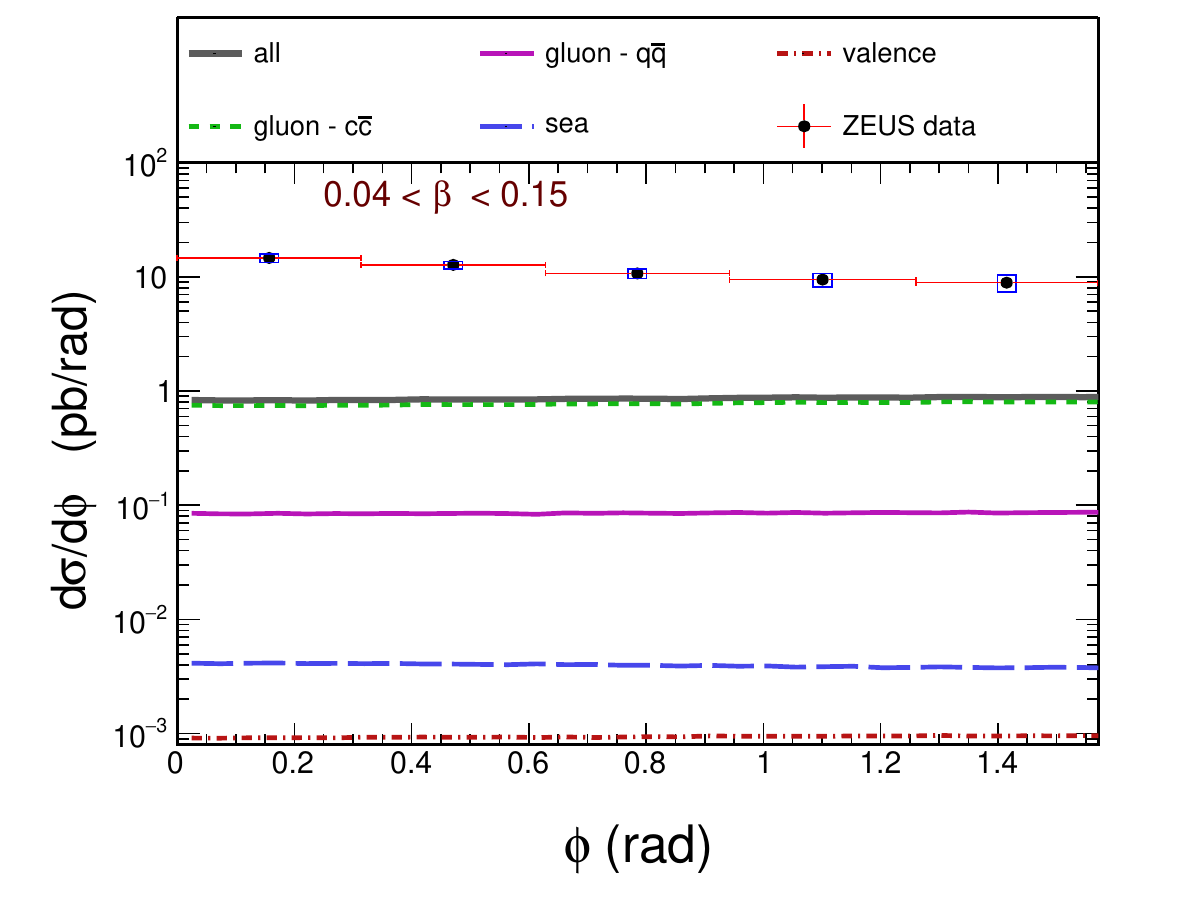}
    \hspace{1em}
    \includegraphics[width=0.45\linewidth]{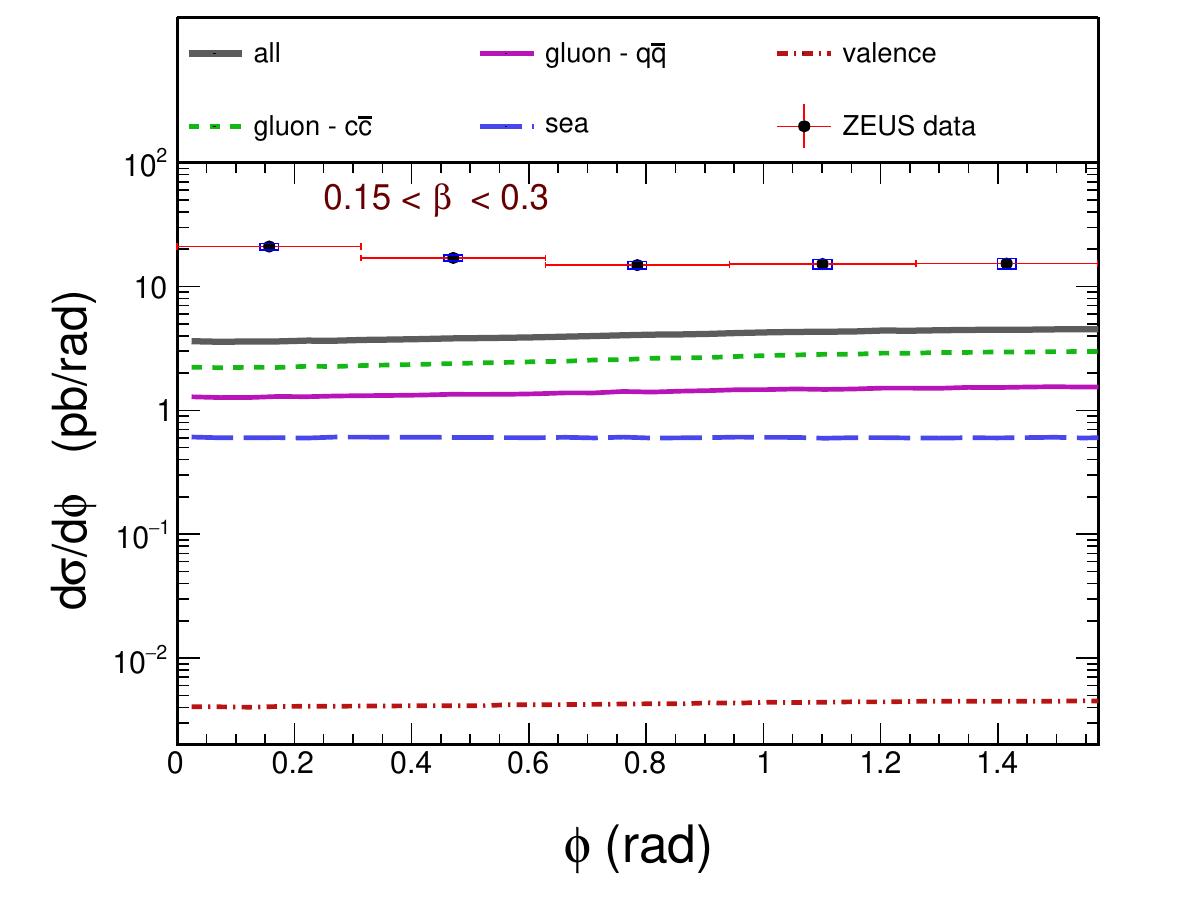}
    \hspace{1em}
    \includegraphics[width=0.45\linewidth]{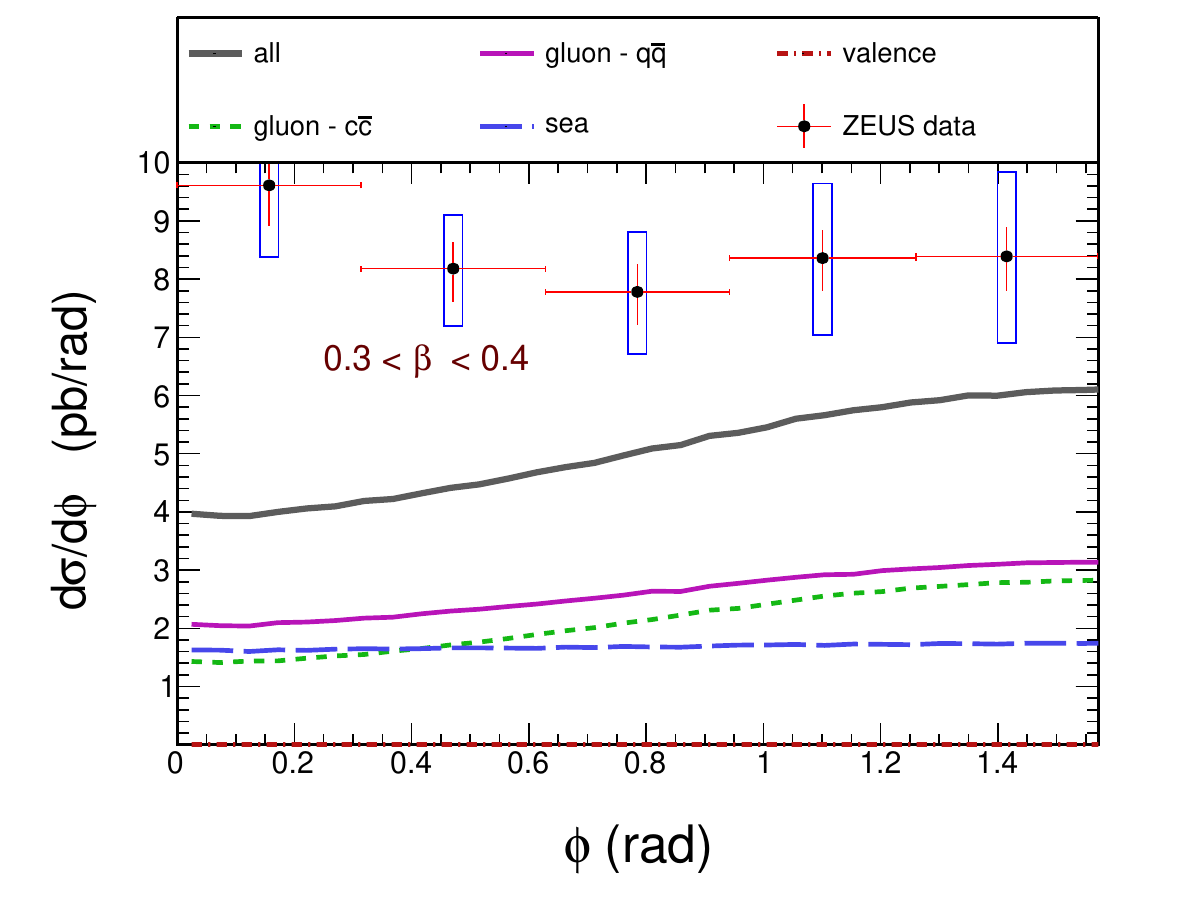}
    \hspace{1em}
    \includegraphics[width=0.45\linewidth]{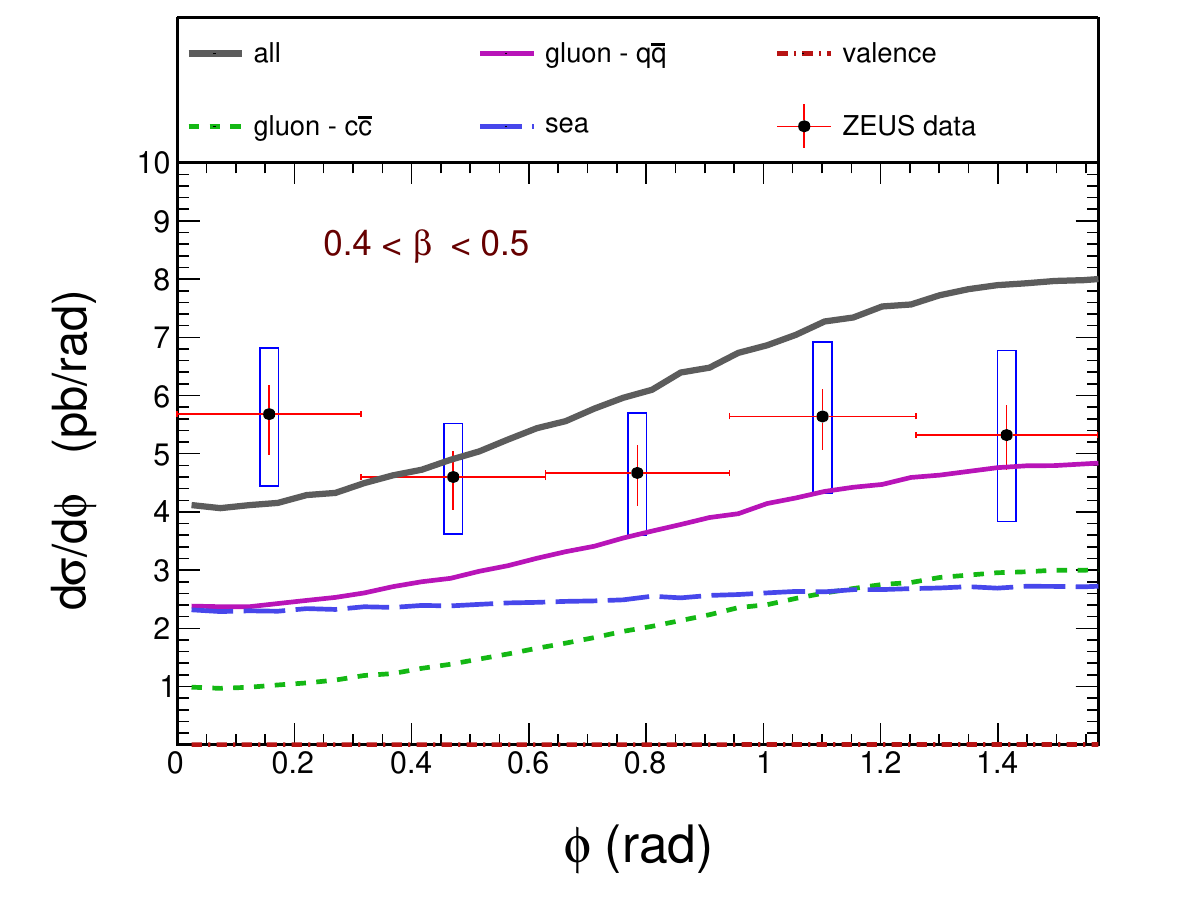}
    \hspace{1em}
    \includegraphics[width=0.45\linewidth]{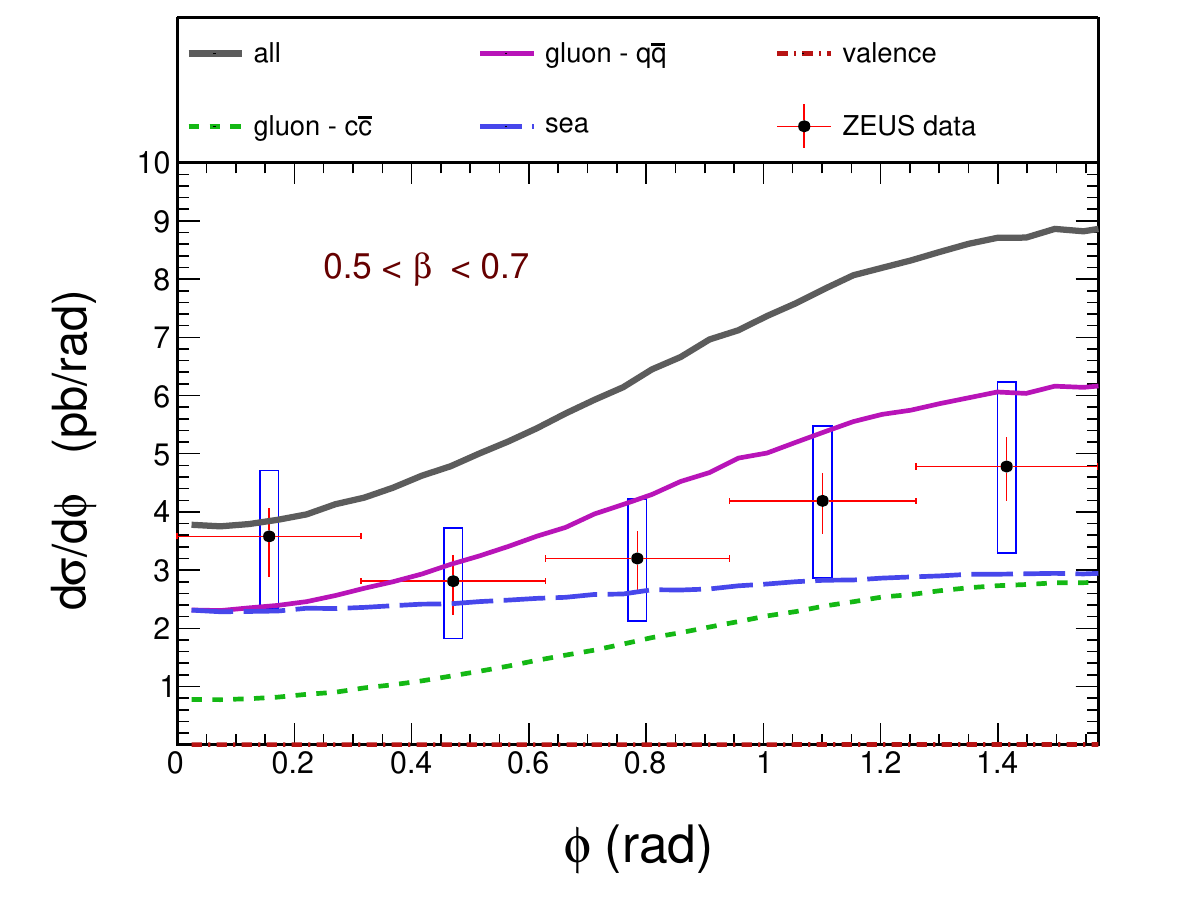}
    \caption{Cross section distributions with the ZEUS cuts imposed along with the available data. The top-left panel shows the differential distribution in the $\beta$ parameter with the ZEUS data shown in black, whereas all the other panels show dependence of the cross section on azimuthal angle $\phi$ with the ZEUS cuts corresponding to five different bins of $\beta$. In these azimuthal distributions, the ZEUS data have been shown with red error bars which indicate the statistical and systematic errors and the blue boxes show the total error range including the normalization uncertainty of the subtraction of the proton-dissociative contribution. Identification of the curves is the same as in Fig.~\ref{fig:Qt}.}
    \label{fig:phi}
\end{figure*}
\begin{figure*}[!htb]
    \centering
    \includegraphics[width=0.45\linewidth]{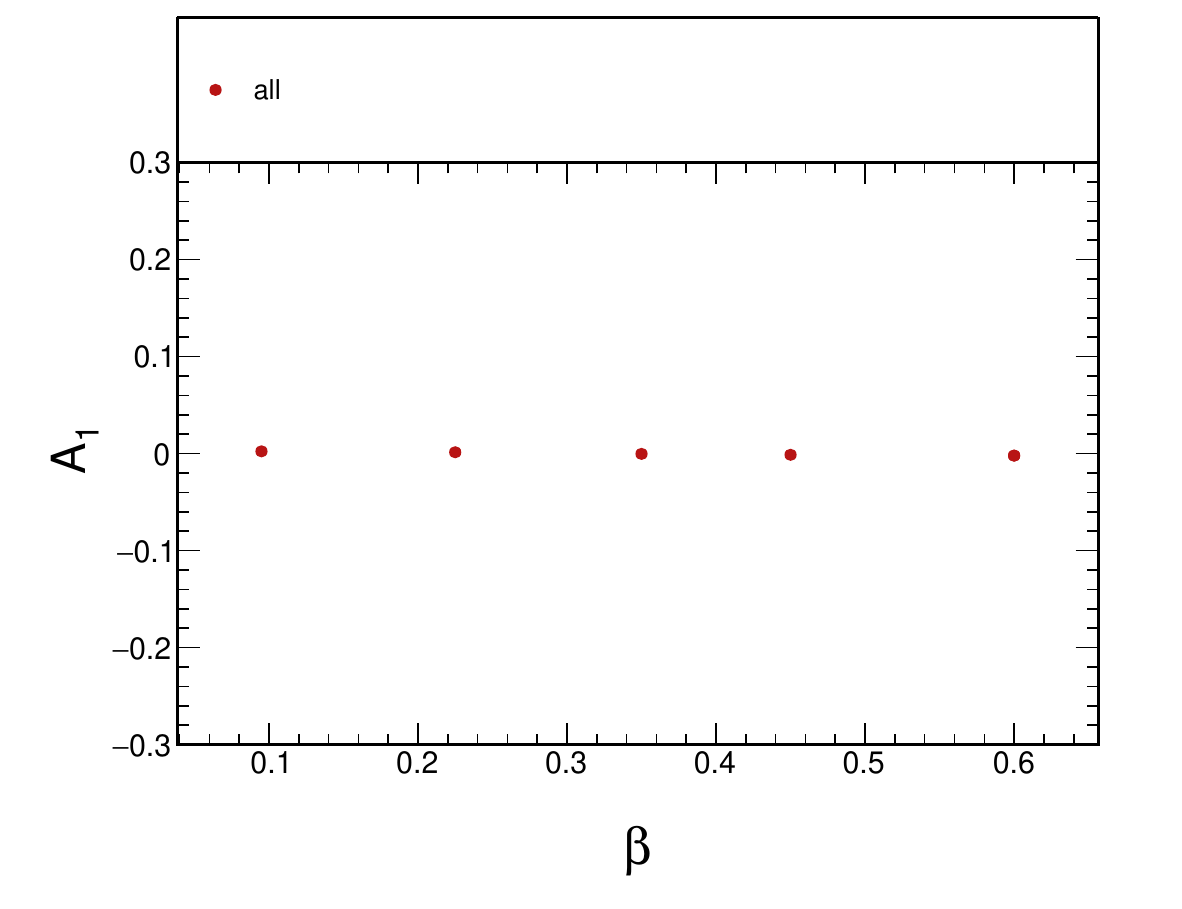} 
    \includegraphics[width=0.45\linewidth]{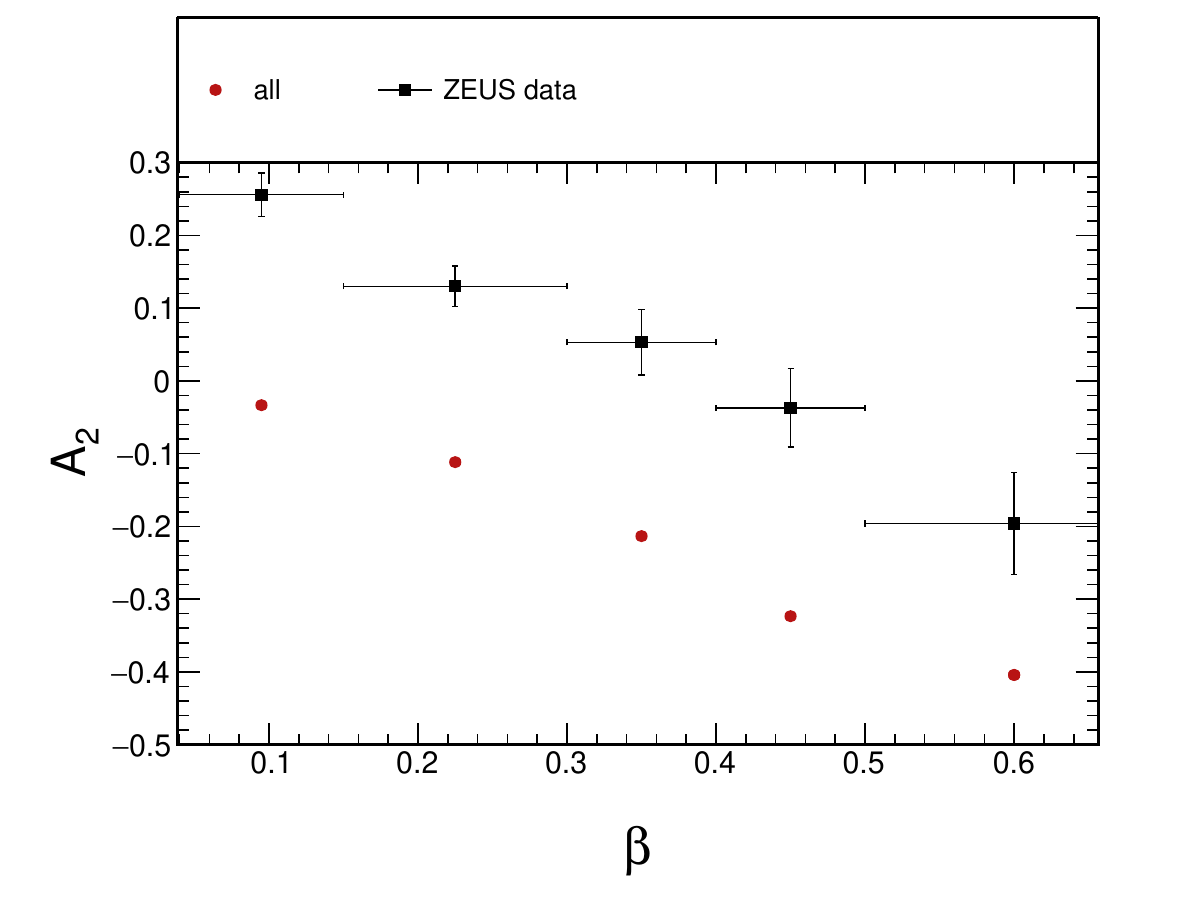}
    \includegraphics[width=0.45\linewidth]{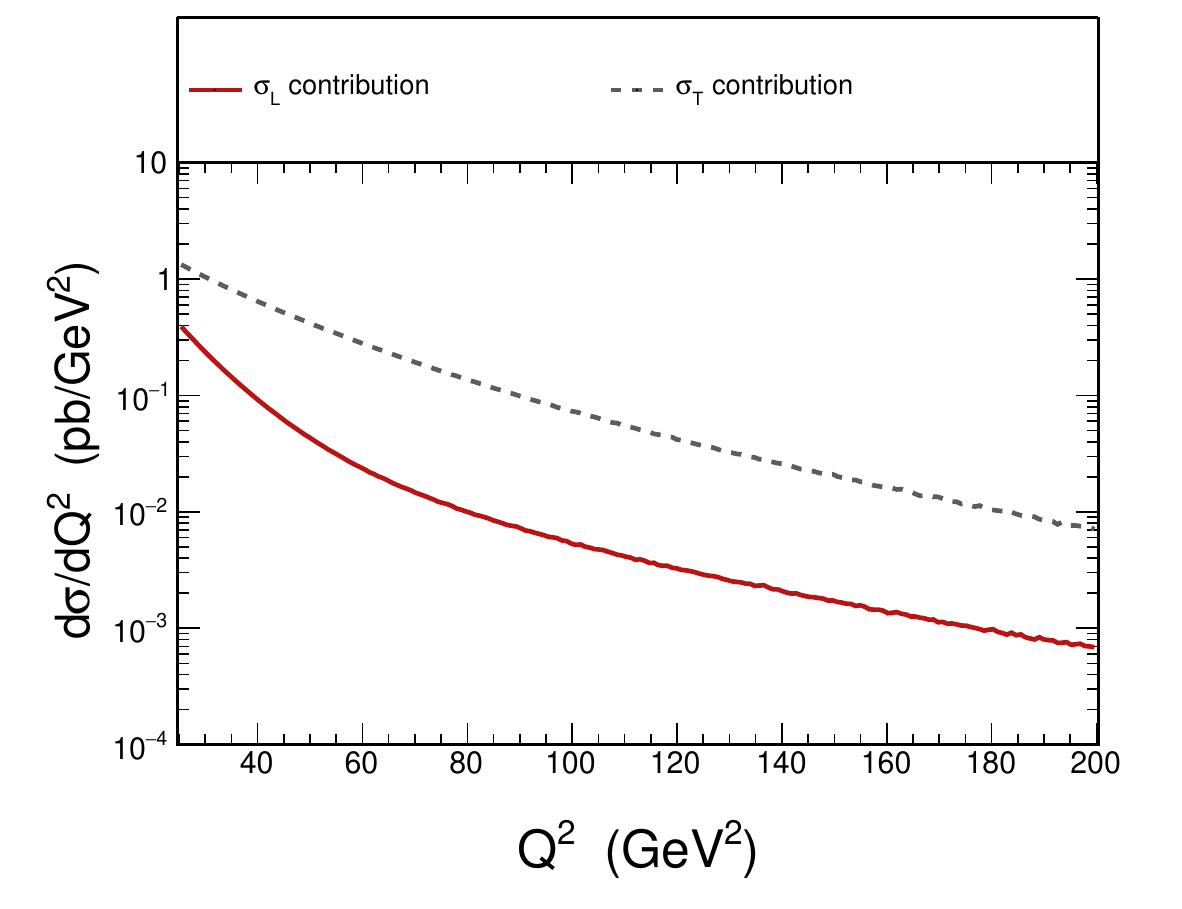}
    \caption{In the top-left and the top-right panel we show the $\beta$-dependence of the Fourier coefficients of the azimuthal distribution, $A_1$ and $A_2$ respectively, in red dots. The ZEUS data are also shown in black in the top-right panel. In the bottom panel, we show both the longitudinal (solid red curve) and the transverse (dashed black curve) contribution to the cross section distribution in $Q^2$ for the ZEUS kinematics. The results include the total contribution across all parts coming from exchanges of valence and sea quarks as well as gluons for light $q\bar{q}$ and heavy $c\bar{c}$ final state.}
    \label{fig:a}
\end{figure*}
In the azimuthal angle distributions shown in Fig.~\ref{fig:phi} corresponding to the ZEUS kinematical region, we have shown the currently available ZEUS data for each of the beta bins: $0.04<\beta<0.15$ in the top-right panel, $0.15<\beta<0.3$ in the middle-left panel, $0.3<\beta<0.4$ in the middle-right panel, $0.4<\beta<0.5$ in the bottom-left panel and $0.5<\beta<0.7$ in the bottom-right panel. In addition, in the top-left panel of Fig.~\ref{fig:phi} we show the distributions in $\beta$ with the ZEUS cuts imposed and with the available data. 
We can clearly see here that when binning at high values of $\beta$, the agreement to data is better than that of lower values of $\beta$. In particular below $\beta \lesssim 0.3$ the $q \bar q$ contribution to dijets falls significantly short of the data. We remind the reader that small $\beta$ corresponds to large invariant masses $M$ of the diffractive system. 

To extract the Fourier coefficients of the $\phi$ distribution in each of the five $\beta$ bins we consider the normalized distribution in $\phi$ to be,
\begin{eqnarray}
    \frac{1}{\sigma }\, \frac{d \sigma}{d \phi} = \frac{1}{2 \pi} \Big( 1 + A_1 \cos \phi + A_2 \cos 2 \phi \Big) \, , \quad \sigma = \int_0^{2 \pi} d \phi \, \frac{d \sigma}{d\phi} \, ,
\end{eqnarray}
Then, the Fourier coefficients $A_1,A_2$ are obtained from
\begin{eqnarray}
    A_1 = \frac{2}{\sigma} \int_0^{2 \pi} d \phi \, \frac{d \sigma}{d\phi} \cos \phi, \quad
    A_2 = \frac{2}{\sigma} \int_0^{2 \pi} d \phi \, \frac{d \sigma}{d\phi} \cos 2 \phi \, 
    .
\end{eqnarray}
The ZEUS collaboration \cite{ZEUS:2015sns} has extracted the coefficient $A_2$ as a function of $\beta$ by fitting the data assuming $A_1 = 0$. We show our results for both $A_1$ and $A_2$ as functions of $\beta$ in the top panel of Fig.~\ref{fig:a}.
Indeed the smallness of $A_1$ is borne out by our calculation. Notice that $A_1$ originates from the interference of longitudinal and transverse virtual photon contributions. While in the ZEUS range $\sigma_L < \sigma_T$ (see the lower panel in Fig.~\ref{fig:a}), the very strong suppression of $A_1$ can be inferred from Eq.~\ref{cs2} for the $LT$--interference, see especially the prefactor $\left(1-2z\right)$ in the gluon contribution.

The coefficient $A_2$ has its origin in the interference of different transverse photon polarizations. 
We compare our results for $A_2$, with the ZEUS data in the top-right panel of Fig.~\ref{fig:a}.
Our calculation yields a negative value of $A_2$ for all $\beta$--bins. While the trend of $A_2$ as a function of $\beta$ is similar as in data, our calculation does not yield the required sign change of $A_2$ at small $\beta$. As to the magnitude of the effect we predict a too large absolute value $A_2$, but it can be expected, that  hadronization effects would dilute the strength of azimuthal correlation. We also note that in \cite{Bartels:1996tc} a negative $A_2$ was argued to be a sign of two-gluon exchange, while the so-called resolved Pomeron contributions resemble one-gluon exchange with the $q \bar q$--system and would give positive $A_2$.

Here one should note that some part of the disagreement with the present HERA dijet data including both H1 and ZEUS can in principle arise also from the uncertainties in the numerical calculation. Similar analyses with other PDF inputs would help to pin down these uncertainties.
\section{Conclusions}
\label{sec:4}
In this work, we have presented the collinear formalism for calculating the exclusive diffractive dijet production relevant for the $e p \to e'jj p$ interaction. 
All leading-order mechanisms have been included: the gluonic mechanism for light and heavy quark pair production, and also contributions from light sea and valence quark exchanges.
Whenever possible, we reproduced the distributions presented in \cite{Braun:2005rg}, a first paper in which the formalism was discussed.
We have presented results of our calculation of different differential distributions that can be studied experimentally.

We have considered both the most general situation (without any experimental cuts imposed) and also the two cases of the H1 \cite{H1:2011kou} and ZEUS \cite{ZEUS:2015sns} experiments imposing cuts corresponding to their specific kinematics.
The aim of the analysis has been to first explore the role of different parton components in different regions of the phase space and then to impose ZEUS kinematical cuts to draw a comparison.
Special attention has been devoted to the valence quark contribution which was not studied so far in the literature.
We discovered that to explore this contribution one would need to go to larger $x_{\mathbb{P}}>0.1$ i.e. beyond the purely diffractive mechanism.
At HERA the main focus was only on the purely diffractive mechanisms.
Perhaps one could explore the mechanisms extended to the nondiffractive regime at the EIC. 

We have calculated the differential distributions relevant for the H1 and ZEUS experiments at HERA, however our theoretical results have no region of agreement with the H1 data, possibly owing to the fact that dijets produced at H1 were not exclusive. So, we have focused on the ZEUS kinematics.
We have discussed how the inclusion of contributions across the various parton exchanges such as gluon (for both the light and heavy quark pair production), sea and valence quarks can affect the observables in the extensive phase space as well as those corresponding to the ZEUS kinematics in particular.
We have shown that adding these new degrees of freedom changes the results only slightly.
We conclude that the new components cannot be responsible for the obvious disagreement of the theoretical results with the H1 data. We have shown the azimuthal correlations for the ZEUS kinematics separately for different windows of diffractive parameter $\beta$ and obtained reasonable agreement in the region of $\beta \gtrsim 0.4$.\\

\section*{Acknowledgments}
This work was supported by the Polish National Science Center Grant No. UMO-2023/49/B/ST2/03665.

\appendix

\section{Integrals}
\label{sec:Integrals}

We introduce the following shorthand notation for integrals over the GPDs (form factors) that depend only on $\xi$ and $\Delta^2$.

\subsection{Gluons}

For gluons, there are two independent integrals:

\begin{eqnarray}
	f^{(1)}_g[H](\xi, \Delta^2) &=& - \int_{-1}^{1}dx\left[ \frac{H^{g}\left(x,\xi,\Delta^2\right)-H^{g}\left(\xi,\xi,\Delta^2\right)}{x- \xi} - \frac{H^{g}\left(x,\xi,\Delta^2\right)-H^{g}\left(-\xi,\xi,\Delta^2\right)}{x+ \xi}\right]  \nonumber \\
	&+&\left [H^{g}\left(-\xi,\xi,\Delta^2\right)\log\left|\frac{1+\xi}{1-\xi}\right|-H^{g}\left(\xi,\xi,\Delta^2\right)\log\left|\frac{1-\xi}{1+\xi}\right|\right] \nonumber \\
	&+& i \pi \left[H^{g}\left(-\xi,\xi,\Delta^2\right)+H^{g}\left(\xi,\xi,\Delta^2\right)\right] \,,
\end{eqnarray}
and 
\begin{eqnarray}
f^{(2)}_g[H](\xi, \Delta^2) &=& \xi \int_{-1}^{1}dx\left[ \frac{H^{g\prime}\left(x,\xi,\Delta^2\right)-H^{g\prime}\left(-\xi,\xi,\Delta^2\right)}{\left(x+ \xi\right)} + \frac{H^{g\prime}\left(x,\xi,\Delta^2\right)-H^{g\prime}\left(\xi,\xi,\Delta^2\right)}{\left(x- \xi\right)}\right]\nonumber\\ &+& \xi \left[H^{g\prime}\left(-\xi,\xi,\Delta^2\right)\log\left|\frac{1+\xi}{1-\xi}\right|+H^{g\prime}\left(\xi,\xi,\Delta^2\right)\log\left|\frac{1-\xi}{1+\xi}\right|\right]\nonumber\\ &+& i \pi \xi \left[H^{g\prime}\left(-\xi,\xi,\Delta^2\right)-H^{g\prime}\left(\xi,\xi,\Delta^2\right)\right] \, . 
\end{eqnarray}

We then have (neglecting contributions with the GPD $E^{g}$)
\begin{eqnarray}
	\mathcal{I}_L^g = \Big \{ (1-2 \beta) f^{(1)}_g[H](\xi, \Delta^2) + 2 (1-\beta) f^{(2)}_g[H](\xi, \Delta^2) \Big \} \frac{\bar u(p') \slashed{n}_{-}u(p)}{2(P \cdot n_{-})}\,,
\end{eqnarray}
and, for transverse polarization,
\begin{eqnarray}
	\mathcal{I}_T^g = \Big \{ - 2\beta f^{(1)}_g[H](\xi, \Delta^2) + (1- 2 \beta) f^{(2)}_g[H](\xi, \Delta^2) \Big \} \frac{\bar u(p') \slashed{n}_{-}u(p)}{2(P \cdot n_{-})}\,.
\end{eqnarray}

\subsection{Quarks}

For quarks, there are four independent integrals:

\begin{eqnarray}
    f_q^{(1)}[H](\xi,\Delta^2) &=& 2\xi \int_{-1}^{1} dx \, \frac{H^q(x,\xi,\Delta^2)}{x - \xi + i \epsilon }\,,   \nonumber \\
    f_q^{(2)}[H](\xi,\Delta^2) &=& 2\xi \int_{-1}^{1} dx \, \frac{H^q(x,\xi,\Delta^2)}{x + \xi - i \epsilon }\,,
    \nonumber \\
    f_q^{(3)}[H](\xi,\beta,\Delta^2) &=& 2 \xi \int_{-1}^{1} dx \, \frac{H^q(x,\xi,\Delta^2)}{x - \xi(1-2\beta)- i \epsilon }\,, \nonumber \\
    f_q^{(4)}[H](\xi,\beta,\Delta^2) &=& 2 \xi \int_{-1}^{1} dx \, \frac{H^q(x,\xi,\Delta^2)}{x + \xi(1-2\beta) + i \epsilon }.
\end{eqnarray}
The above integrals can be deduced as
\begin{eqnarray}
    f_q^{(1)}[H]\left(\xi,\Delta^2\right) &=& 2\xi\Bigg[\int_{-1}^{1} dx \, \frac{H^q\left(x,\xi,\Delta^2\right)-H^q\left(\xi,\xi,\Delta^2\right)}{x - \xi }-i\pi H^q\left(\xi,\xi,\Delta^2\right)\nonumber\\ &+& H^q\left(\xi,\xi,\Delta^2\right)\log{\left|\frac{1-\xi}{1+\xi}\right|} \Bigg]\,, \nonumber \\
    f_q^{(2)}[H]\left(\xi,\Delta^2\right) &=& 2\xi\Bigg[\int_{-1}^{1} dx \, \frac{H^q\left(x,\xi,\Delta^2\right)-H^q\left(-\xi,\xi,\Delta^2\right)}{x + \xi }+i\pi H^q\left(-\xi,\xi,\Delta^2\right)\nonumber\\ &+& H^q\left(-\xi,\xi,\Delta^2\right)\log{\left|\frac{1+\xi}{1-\xi}\right|} \Bigg]\,, \nonumber 
    \nonumber \\
    f_q^{(3)}[H]\left(\xi,\beta,\Delta^2\right) &=& 2 \xi \Bigg[\int_{-1}^{1} dx \, \frac{H^q\left(x,\xi,\Delta^2\right)-H^q\left(\xi\left(1-2\beta\right),\xi,\Delta^2\right)}{x - \xi(1-2\beta)}+ i\pi H^q\left(\xi\left(1-2\beta\right),\xi,\Delta^2\right)\nonumber\\ &+& H^q\left(\xi\left(1-2\beta\right),\xi,\Delta^2\right)\log{\left|\frac{1-\xi\left(1-2\beta\right)}{1+\xi\left(1-2\beta\right)}\right|}\Bigg]\,, \nonumber \\
    f_q^{(4)}[H]\left(\xi,\beta,\Delta^2\right) &=&  2 \xi
    \Bigg[\int_{-1}^{1} dx \, \frac{H^q\left(x,\xi,\Delta^2\right)-H^q\left(-\xi\left(1-2\beta\right),\xi,\Delta^2\right)}{x + \xi(1-2\beta)}- i\pi H^q\left(-\xi\left(1-2\beta\right),\xi,\Delta^2\right)\nonumber\\ &+& H^q\left(-\xi\left(1-2\beta\right),\xi,\Delta^2\right)\log{\left|\frac{1+\xi\left(1-2\beta\right)}{1-\xi\left(1-2\beta\right)}\right|} \Bigg]. 
    \label{a6}
\end{eqnarray}
We then have that (neglecting contributions with the GPD $E^{q}$)
\begin{eqnarray}
    \mathcal{I}_L^q &=&\Big\{ zf_q^{(1)}[H](\xi,\Delta^2)+\bar{z}f_q^{(2)}[H]\left(\xi,\Delta^2\right)\Big\} \frac{\bar u(p') \slashed{n}_{-}u(p)}{2(P \cdot n_{-})},
    \label{a7}
\end{eqnarray}    
and, for the transverse polarization,
\begin{eqnarray}
    \mathcal{I}_T^{q_1} &=& \Big\{z\bar{z}f_q^{(1)}[H](\xi,\Delta^2) -\frac{\beta \bar{z}^2}{\bar{\beta}}f_q^{(2)}[H](\xi,\Delta^2) + \frac{\bar{z}^2}{\bar{\beta}} f_q^{(3)}[H](\xi,\beta,\Delta^2)\Big\} \frac{\bar u(p') \slashed{n}_{-}u(p)}{2(P \cdot n_{-})}\,,\nonumber\\ 
    \mathcal{I}_T^{q_2} &=&\Big\{ -z\bar{z}f_q^{(2)}[H](\xi,\Delta^2) +\frac{\beta z^2}{\bar{\beta}}f_q^{(1)}[H](\xi,\Delta^2) - \frac{z^2}{\bar{\beta}} f_q^{(4)}[H](\xi,\beta,\Delta^2)\Big\}\frac{\bar u(p') \slashed{n}_{-}u(p)}{2(P \cdot n_{-})}\,, \nonumber \\
\end{eqnarray}
where $\bar{z}=1-z$ and $\bar{\beta}=1-\beta$.
\bibliography{dijet.bib}

\end{document}